\documentclass{jfm}
\usepackage{subfigure,upmath,bm,amsmath,amssymb,natbib,graphicx}
\usepackage{array,epic,eepic,color,verbatim,indentfirst}
\pdfoutput=1

\newcommand{\MRW}{\mbox{\it MRW}} 
\newcommand{\MSW}{\mbox{\it MSW}} 

\newlength{\pacolength}

\title[Transitions to 3D flows in a cylinder driven by
  oscillations of the sidewall] 
{Transitions to three-dimensional flows in a cylinder driven by
  oscillations of the sidewall} 

\author[C. Panades, F. Marques and J.\ M. Lopez]{C.\ns P\ls A\ls N\ls
  A\ls D\ls E\ls S,$^1$ F.\ns M\ls A\ls R\ls Q\ls U\ls E\ls S$^1$ and
  J.\ns M.\ns L\ls O\ls P\ls E\ls Z$^2$ }

\affiliation{
 $^1$Departament de F{\'\i}sica Aplicada, Universitat Polit\`ecnica de
  Catalunya, Barcelona 08034, Spain\\[\affilskip]

 $^2$School of Mathematical and Statistical Sciences, Arizona State
 University, Tempe AZ 85287, USA}

\pubyear{} \volume{} \pagerange{}
\date{3 November 2010}
\begin{document}

\maketitle

\begin{abstract}

The transition from two-dimensional to three-dimensional flows in a
finite circular cylinder driven by an axially oscillating sidewall is
explored in detail. The complete symmetry group of this flow,
including a spatio-temporal symmetry related to the oscillating
sidewall, is $Z_2\times O(2)$. Previous studies in flows with the same
symmetries, such as symmetric bluff-body wakes and periodically forced
rectangular cavities, were unable to obtain the theoretically
predicted bifurcation to modulated traveling waves. In the simpler
cylindrical geometry, where the azimuthal direction is physically
periodic, we have found these predicted modulated traveling waves as
stable fully saturated nonlinear solutions for the first time. A
careful analysis of the base states and their linear stability
identifies different parameter regimes where three-dimensional states
that are either synchronous with the forcing or quasiperiodic,
corresponding to different symmetry-breaking processes. These results
are in good agreement with theoretical predictions and previous
results in similar flows. These different regimes are separated by
three codimension-two bifurcation points that are yet to be fully
analyzed theoretically.  Finally, the saturated nonlinear states and
their properties in different parameter regimes are analyzed.

\end{abstract}

\section{Introduction}

When a system is invariant under the action of a group of symmetries,
there can be far-reaching consequences on its bifurcations. When the
symmetries are purely spatial in nature (e.g.\ reflections,
translations, rotations), these consequences have been extensively
studied \citep*[e.g., see][]{GoSc85, GSS88, CrKn91, CrHo93, ChIo94,
  IoAd98, ChLa00, GoSt02}. The system may also be invariant to the
action of spatio-temporal symmetries. These are spatial symmetries
composed with temporal evolution. A classic example is the
two-dimensional K\'arm\'an vortex street form of the wake of a
circular cylinder. Other common cases are periodically forced flows.

The transition from two-dimensional to three-dimensional flow is of
fundamental interest in fluid dynamics. Two-dimensional flows, like
the K\'arm\'an vortex street and other bluff-body wakes, are invariant
in the spanwise direction to both translations ($SO(2)$ symmetry
group) and reflections ($Z_2$ symmetry group), the combination
generating the $O(2)$ symmetry group. We are interested in the
transition from two-dimensional to three-dimensional flow when the
two-dimensional problem has a spatio-temporal symmetry of $Z_2$ type,
as is the case for the wake of a circular cylinder in the streamwise
direction \citep*{BlLo03_PoF, BML05}. Another flow with the same
spatio-temporal symmetries as the periodically shedding wake is that
in a periodically-driven rectangular cavity of infinite spanwise
extent \citep*{MLB04}, which has been studied in \citet*{LoHi01,
  VHL03, BlLo03_jfm}.

The complete symmetry group of these flows is $Z_2\times O(2)$. The
implications of $O(2)$ symmetry in fluid systems have been studied
extensively, both when the instability breaking $O(2)$ symmetry
(i.e.\ transition from two-dimensional to three-dimensional) is due to
a single real eigenvalue becoming positive (steady bifurcation) as
well as when it is due to a pair of complex-conjugate eigenvalues
gaining positive real part, leading to time-periodic flow (e.g.\ see
the references cited above). The types of symmetry-breaking
bifurcations to three-dimensional flow that a two-dimensional flow
with a space-time symmetry can experience are completely determined by
the symmetry group of the system, and not by the particulars of the
physical mechanisms responsible for the bifurcation, and have been
analyzed in detail in \citet{MLB04}. The main results obtained are
that there are two type of bifurcations, one synchronous with the
forcing and the other resulting in quasiperiodic flows. Both types
come in two different flavors, depending on the symmetries of the
bifurcated solutions. There are two synchronous modes, A and B, that
break or preserve the space-time symmetry $Z_2$, respectively. The
quasiperiodic solutions have the form of modulated traveling waves or
modulated standing waves in the spanwise direction; they differ in
their symmetry properties: the traveling waves preserve a space-time
symmetry, while the standing wave preserves a purely spatial reflection
symmetry.

In the examples of flows with $Z_2\times O(2)$ symmetry group
described above, the $O(2)$ invariance is only an idealization of the
corresponding experimental flow due to the finite extent of the
spanwise direction. The typical result is that the travelling waves
predicted by the theory do not travel, due to endwall effects
\citep{LHBML05}.  Cylindrical geometries are very useful in the sense
that the azimuthal direction is physically periodic and have the
$O(2)$ symmetry group exactly fulfilled. This fact has been explored
in \citet{BlLo10} in a driven annular geometry, but unfortunately, the
modulated traveling wave modes that are predicted from the Floquet
analysis do not saturate nonlinearly to pure modes but are always
mixed with contributions from the synchronous A mode. In the present
paper we explore a simpler setting, a finite circular cylinder with an
axially oscillating sidewall. The base state is also $Z_2\times O(2)$
invariant as in the other flow examples, and we have indeed found
nonlinearly saturated pure modulated travelling waves for the first
time in a physically-realizable flow. In this cylindrical setting, as
the travelling waves move in the azimuthal direction, i.e.\ the
pattern rotates around the cylinder axis, they will be termed rotating
(or modulated rotating) waves.

The paper is organized as follows: in \S\ref{gov_equat} the
formulation of the problem and numerical methods used are presented;
in \S\ref{base_states} the base state of the system is computed, and
its changes when parameters are varied are discussed; in
\S\ref{stability} the linear stability of the basic flow is studied,
and compared with similar flows. In \S\ref{3Dmodes} the
three-dimensional structure and symmetries of the different unstable
modes found are analyzed in detail. Finally, in \S\ref{conclusions} the
main results are summarized and open problems and future directions of
study are discussed.

\section{Governing equations and numerical methods}\label{gov_equat}

\begin{figure}
\begin{center}
\includegraphics[width=0.4\linewidth]{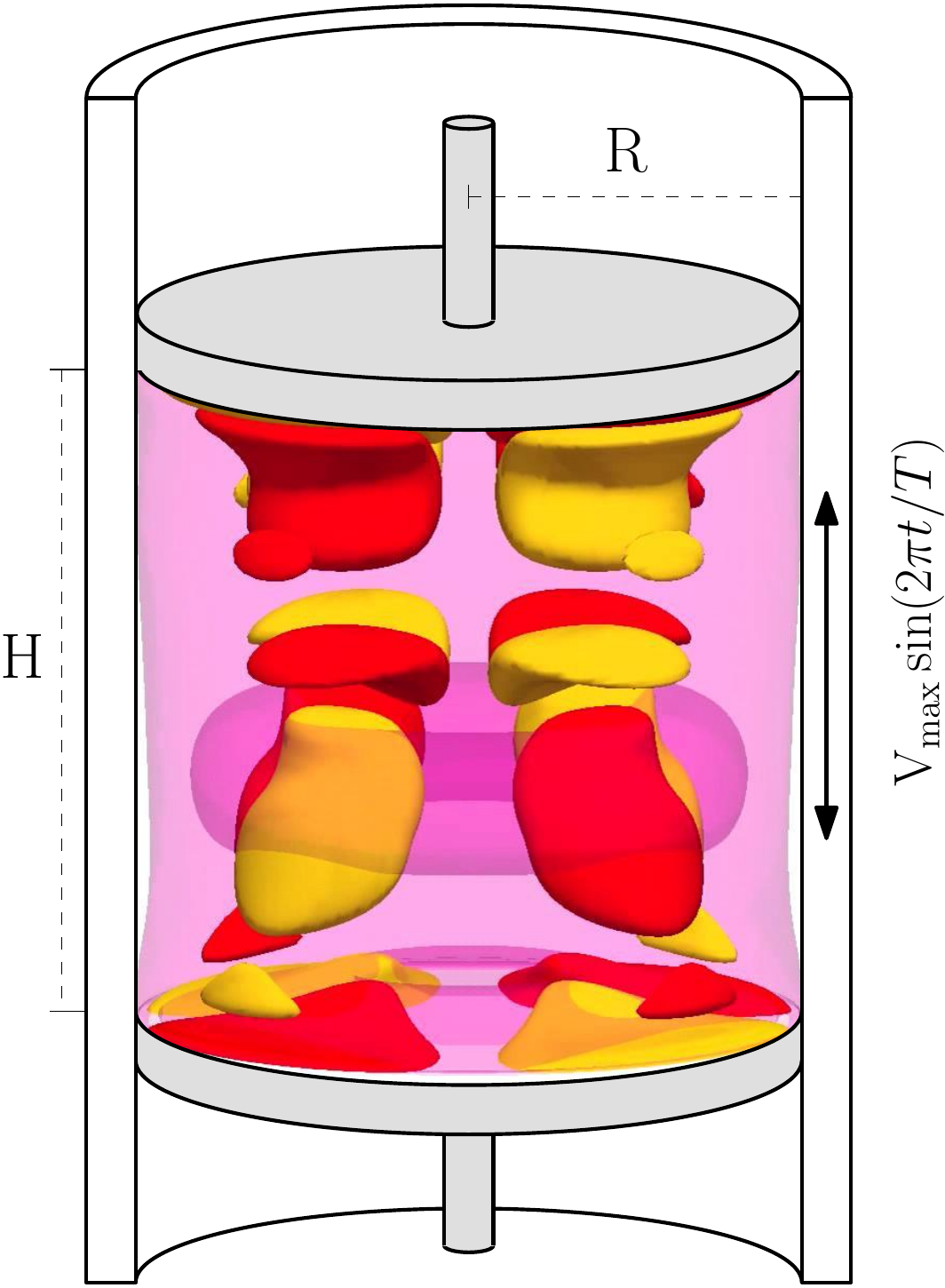}
\end{center}
\caption{(Color online) Schematic of the apparatus. The inset shows a
  synchronous bifurcated state; translucent isosurfaces (of azimuthal
  vorticity) show the axisymmetric roller generated by the side wall
  oscillation, and solid isosurfaces (of radial vorticity) show the
  braid structures associated with three-dimensional instabilities.}
\label{schematics}
\end{figure}

Consider a Newtonian fluid of kinematic viscosity $\nu$ confined in
a finite cylinder of radius $R$ and height $H$, whose sidewall
oscillates harmonically in the axial direction, with period $T$ and
maximum axial velocity $V_{\max}$, while the top and bottom lids remain
at rest, as shown schematically in figure~\ref{schematics}. The
system is non-dimensionalized taking $R$ as the length scale, and the viscous
time $R^2/\nu$ as the time scale. There are three non-dimensional
parameters in this problem:
\begin{alignat}{2}
  & \text{Aspect ratio}    &\qquad& \varGamma=H/R,    \\
  & \text{Reynolds number} &&       Re=V_{\max} R/\nu, \\
  & \text{Stokes number}   &&       St=R^2 /\nu T.
\end{alignat}
The aspect ratio defines the geometry of the problem, while $Re$ and
$St$ are non-dimensional measures of the amplitude and frequency of the
forcing; the inverse of the Stokes number is precisely the
non-dimensional period of the oscillations, $\tau=1/St$. In the
current study, the aspect ratio is fixed at $\varGamma=2$.
The non-dimensional Navier--Stokes equations governing the flow are
\begin{subequations}
  \begin{gather}
    \Big(\frac{\partial}{\partial t}+\mathbf{u}\cdot\nabla\Big)\mathbf{u}=
      -\nabla p+\nabla^2\mathbf{u}, \\
    \nabla\cdot\mathbf{u}=0,
  \end{gather}
\end{subequations}
where $\mathbf{u}=(u,v,w)$ is the velocity field in cylindrical
coordinates $(r,\theta,z) \in[0,1]\times[0,2\pi]\times
[-\varGamma/2,\varGamma/2]$, and $p$ is the kinematic pressure, all
non-dimensional. The vorticity associated to the velocity field is
$\nabla \times \mathbf{u}=(\xi, \eta, \zeta)$. No-slip
velocity boundary conditions are used on all walls. The velocity
is zero on stationary top and bottom endwalls, and the $z$-component
of velocity at the sidewall
oscillates periodically in time:
\begin{subequations}
  \begin{align}
    & \mathbf{u}(r,\theta,\pm\varGamma/2,t)= (0,0,0), \\
    & \mathbf{u}(1,\theta,z,t)=\big(0,0,Re\sin(2\pi St\, t)\big).
      \label{BCw}
  \end{align}
\end{subequations}
These idealized boundary conditions are discontinuous at the junctions
where the stationary lids meet the oscillating sidewall. In a physical
experiment there are small but finite gaps at these junctions where
the axial velocity adjusts rapidly to zero. For a proper use of
spectral techniques, a regularization of this discontinuity is
implemented of the form
\begin{equation}
 w(1,\theta,z,t)= Re\sin(2\pi St\, t)
 \left[1-\exp{\left(-\frac{1-z}{\epsilon}\right)} \right]
 \left[1-\exp{\left(-\frac{1+z}{\epsilon}\right)} \right]
\end{equation}
where $\epsilon$ is a small parameter that mimics the small physical
gaps ($\epsilon=6 \times 10^{-3}$ has been used as a fixed
parameter). The use of $\epsilon \neq 0$ regularizes the otherwise
discontinuous boundary conditions; see \citet{LoSh98} for further
details on the use of this technique in spectral codes.

Instead of having a oscillatory sidewall, we could also consider the
situation where the sidewall is at rest, and the two endwalls
oscillate harmonically, which in some cases can be more convenient
from the experimental point of view. In order to have a fixed domain,
it is very useful to write the governing equations in the oscillating
reference frame, in which the cylindrical domain is at rest and the
sidewall oscillates. However, the oscillating reference frame is not
an inertial frame, and inertial body force terms must be included in
the Navier--Stokes equations. In this case, there is one extra term,
$-{\bf A}/\rho$, where ${\bf A}$ is the acceleration of the lids and
$\rho$ is the fluid density. Since the acceleration ${\bf A}$ depends
only on time, and for an incompressible flow $\rho$ is constant, this
extra term is a gradient that can be incorporated in the pressure
term, thereby recovering exactly the same formulation as in the case
of oscillatory sidewalls. The difference between oscillating the
sidewall or the endwalls is only important when density variations
appear, for example in the presence of temperature differences, or
concentration differences if the fluid is a mixture.

The governing equations and boundary conditions are invariant to the
following spatial symmetries:
\begin{subequations}
 \begin{equation}
  K_\theta (u,v,w)(r,\theta,z,t)=(u,-v,w)(r,-\theta,z,t),
 \end{equation}
 \begin{equation}
  R_\alpha (u,v,w)(r,\theta,z,t)=(u,v,w)(r,\theta+\alpha,z,t),
 \end{equation}
\end{subequations}
for any real $\alpha$. $K_\theta$ represents reflections about any
meridional plane, whilst $R_\alpha$ signifies rotations about the
cylinder axis. $K_\theta$ and $R_\alpha$ generate the groups $Z_2$ and
$SO(2)$, but the two operators do not commute, so the symmetry group
generated by $K_\theta$ and $R_\alpha$ is $O(2)$ and it acts in the
periodic azimuthal $\theta$-direction. The horizontal reflection on
the mid-plane $z=0$ acts on the velocity field as:
\begin{equation}
 K_z(u,v,w)(r,\theta,z,t)=(u,v,-w)(r,\theta,-z,t).
\end{equation}
However, $K_{z}$ is not a symmetry of the system: due to the harmonic
oscillation of the sidewall, the boundary condition \eqref{BCw} is not
$K_{z}$ invariant. However, $K_zw(1,\theta,z,t)$ coincides with
$w(1,\theta,z,t+T/2)$, introducing an additional spatio-temporal
symmetry. Therefore, the system is also invariant to a reflection
about the half-height plane $z=0$ together with a half-period
evolution in time:
\begin{equation}
  H(u,v,w)(r,\theta,z,t)=(u,v,-w)(r,\theta,-z,t+\tau/2).
\end{equation}
The transformation $H$ generates another $Z_2$ symmetry group that
commutes with $O(2)$. Hence, the complete symmetry group of the
problem is $Z_2\times O(2)$. The action of the spatio-temporal
symmetry $H$ on the vorticity is different to the action on the
velocity, and is given by:
\begin{equation}
  H\left( \xi, \eta, \zeta \right)(r,\theta,z,t)=\left( -\xi, -\eta,
  \zeta \right)(r,\theta,-z,t+\tau/2).
\end{equation}
Therefore, the individual symmetries (and the generated groups) are
exactly the same as for the periodically-driven annular cavity and
analogous to the two-dimensional time-periodic wake of symmetrical
bluff bodies and periodically-driven rectangular cavity flows.

\subsection{Numerical formulation}\label{num_formul}

The governing equations have been solved using a second-order
time-splitting method. The spatial discretization is via a
Galerkin-Fourier expansion in $\theta$ and a Chebyshev collocation in
$x=2r$ and $z$, of the form
\begin{equation}
 F(r,\theta,z)= \sum_{l=0}^{n_r} \sum_{n=0}^{n_z} \sum_{m=-n_\theta}^{n_\theta}
 a_{l,n,m}T_l(x)T_n(z)e^{im\theta}
\end{equation}
for the three velocity components and pressure. The results presented
here were computed with $n_r=48$, $n_z=96$ and $n_\theta=10$. This
resolution resolves all the spatial scales in the solutions presented
here. Time steps of $\delta t=10^{-5}$ have been required to ensure
numerical stability and accuracy of the temporal scheme. For each
Fourier mode, the corresponding Helmholtz and Poisson equations are
solved efficiently using a complete diagonalization of the operators
in both the radial and axial directions. Note that to decouple the
Helmholtz equations for $u$ and $v$, we have used the combinations
$u_+=u+iv$ and $u_-=u-iv$. The coordinate singularity at the axis
$(r=0)$ is treated following some prescriptions, which guarantee the
regularity conditions at the origin needed to solve the Helmholtz
equations \citep*[see][for details]{MNF91}. The spectral collocation
solver used here is based on a previous scheme \citep*{MBA10} that has
recently been tested and used in a wide variety of flows in enclosed
cylinders \citep{MMBL07,LMMB07,LMRA09,LoMa09}.

\section{Basic states}\label{base_states}

\begin{figure}
\begin{center}
\begin{tabular}{ccccc}
 & $(a)$ $St=10$ & $(b)$ $St=32$ & $(c)$ $St=50$ & $(d)$ $St=100$ \\
\raisebox{0.25\linewidth}{$\psi$} &
\includegraphics[width=0.23\linewidth]{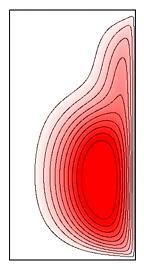} &
\includegraphics[width=0.23\linewidth]{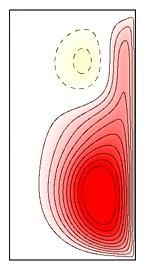} &
\includegraphics[width=0.23\linewidth]{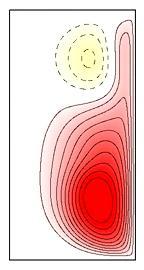} &
\includegraphics[width=0.23\linewidth]{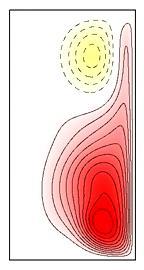}\\
\raisebox{0.25\linewidth}{$\eta$} &
\includegraphics[width=0.23\linewidth]{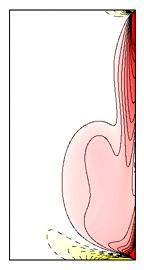} &
\includegraphics[width=0.23\linewidth]{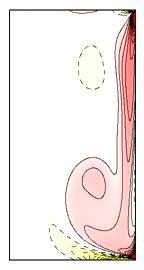} &
\includegraphics[width=0.23\linewidth]{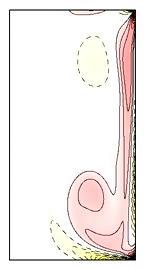} &
\includegraphics[width=0.23\linewidth]{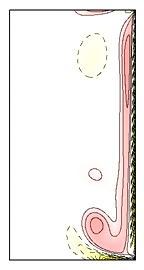}\\
\end{tabular}
\end{center}
\caption{(Color online) Contours of streamfunction, $\psi$, and
  azimuthal vorticity, $\eta$, of the basic state at four $St$ values
  as indicated, for amplitudes $Re$ very close to the corresponding
  critical values. Solid (dashed) contours are positive (negative);
  light/dark (yellow/red online) colors correspond to
  negative/positive values. The base state is periodic, and we have
  selected for each $St$ the phase of period where the oblique jet at
  the bottom corner is most intense; the associated movies online show
  temporal evolution over one period.}
\label{2Dmovies}
\end{figure}

The basic flow, having the symmetries of the problem, is always
axisymmetric and time-periodic, synchronous with the forcing.  The
axial oscillations of the cylindrical wall produce periodic
Stokes-type boundary layers on the oscillating wall. These layers
separate from the sidewall and move towards the cylinder axis after
colliding with the endwalls to form rollers. These rollers are formed
every half period alternatively on each endwall. The term roller
refers to large-scale rotating flow structures with primarily
azimuthal vorticity, $\eta$. Instantaneous contours of the
streamfunction ($\psi$, such that $u=-1/r\, \partial \psi/\partial z$
and $w=1/r\, \partial \psi/\partial r$) are shown in the first row of
figure~\ref{2Dmovies} for four increasing values of the forcing
frequency $St$, and for amplitudes $Re$ very close to, and above, the
critical value at which the basic flow becomes unstable; the online
version of the paper includes movies animating these contours over a
forcing period. In all cases the figures represent meridional planes
$(r,z) \in [0,r] \times [-\varGamma/2,\varGamma/2]$ and the wall is at
rest at $t=0$.

The magnitude and size of the rollers changes substantially with $St$.
For small forcing frequencies, there is sufficient time for these
rollers to dissipate during part of the forcing period, and so in
figure~\ref{2Dmovies}$(a)$ a single roller fills the whole domain most
of the time, whereas for large frequencies the rollers persist at both
ends throughout the whole forcing cycle. The Stokes number determines
the size of the rollers and their dissipation, and the Reynolds number
measures the strength of their collision with the lids and the
recirculation of the fluid. The characteristics of the rollers are
similar to the ones described in previous works for the planar case
\citep{BlLo03_jfm} and for an annular cavity \citep{BlLo10}, but in
the present analysis the curvature effects are very important, and
the flow geometry is substantially altered near the cylinder axis.

Instantaneous contours of the azimuthal vorticity are shown in the
second row of figure~\ref{2Dmovies}. These contours describe the
boundary layers that form at the sidewall and endwalls and their
dynamics very well. The sidewall boundary layer is a Stokes-type
boundary layer whose thickness is proportional to $St^{-1/2}$
\citep{ScKe79,MaLo97}, so the boundary layer becomes thinner for
larger values of the forcing frequency $St$ (it also becomes thinner
as the amplitude of the forcing $Re$ is increased). The sidewall
boundary layer, dragged by the cylinder sidewall motion, separates
upon colliding with the endwalls, and from the corners where the
sidewall and endwalls meet, the boundary layer enters the bulk of the
fluid, forming axisymmetric oblique jets that result in the formation
of the rollers. This process is analogous to the formation of a vortex
roller near the junction of an impulsively started plate and a
stationary plate, where there is a jump in the velocity
\citep{AlLo07}. The jets are clearly seen in the azimuthal vorticity
contours: the jet centerline coincides with the azimuthal vorticity
zero contour, and on each side it is surrounded by regions with
intense azimuthal vorticity of opposite signs.  Oscillating boundary
layers also form on the endwalls with azimuthal vorticity of opposite
sign to that of the rollers, as a result of the jet dynamics just
described, and because the endwalls are at rest.

\section{Stability of the basic flow}\label{stability}

By increasing the amplitude of the forcing $Re$ beyond a critical
value $Re_\textrm{c}(St)$, the basic state undergoes a symmetry-breaking
bifurcation yielding a new three-dimensional state. Depending on $St$,
the basic state may undergo either synchronous or Neimark--Sacker
bifurcations. The stability of the basic flow has been comprehensively
explored for $St \in [1,150]$, revealing two synchronous modes ($A$
and $B$) that bifurcate from the axisymmetric state by breaking the
symmetries differently in each case. There is also a novel
quasiperiodic mode that manifests as modulated rotating waves
\MRW. We use subscripts for each of these states to indicate their
azimuthal wavenumber $m$.

The linear stability of the basic state to general three-dimensional
perturbations has been determined using global linear stability
analysis via time evolution of the Navier--Stokes equations
\citep*{LMRA09,DLM10}. First, a periodic axisymmetric basic state was
computed at some point in parameter space. Its stability was
determined by introducing small random perturbations into all
azimuthal Fourier modes. For sufficiently small perturbations, the
nonlinear couplings between Fourier modes are negligible (below
round-off numerical noise) and the growth rates (real parts of the
eigenvalues) and structure of the eigenfunctions corresponding to the
fastest growing perturbation at each Fourier mode emerge from time
evolution. This is tantamount to a matrix-free generalized power
method in which the actions of the Jacobian matrices for the
perturbations are given by time integration of the Navier--Stokes
equations with the aforementioned initial conditions. This direct
numerical technique is very efficient as the exponential growth or
decay of the perturbations is established in a relatively short
evolution time, and there is no need to evolve the disturbances until
they saturate nonlinearly.

\begin{figure}
\begin{center}
\includegraphics[width=0.7\linewidth]{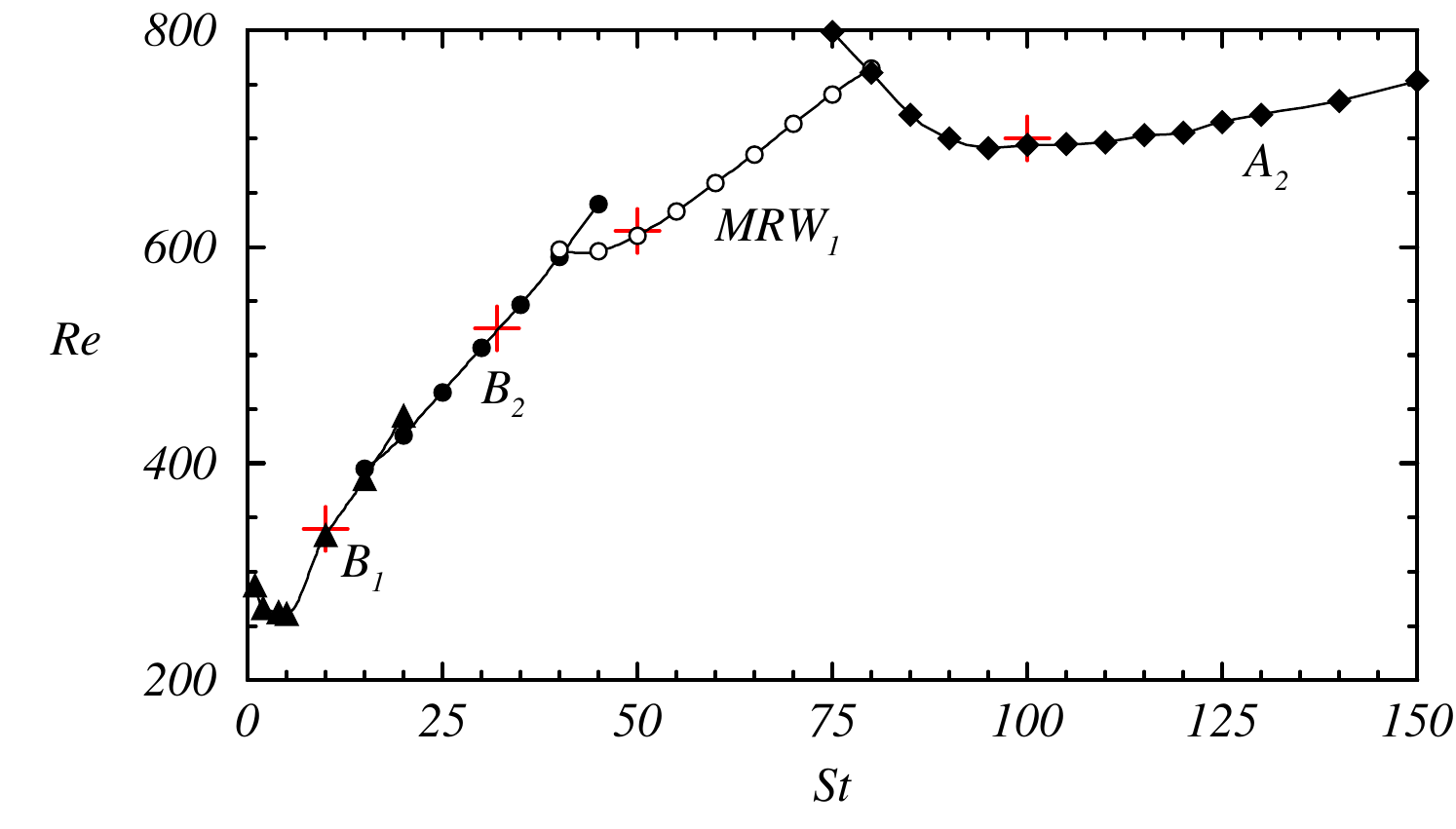}
\end{center}
\caption{(Color online) Critical Reynolds number, $Re_\textrm{c}$, as
  a function of the Stokes number, $St$, for the transition from the
  basic state to the different three-dimensional states, $B_1$, $B_2$,
  $\MRW_1$ and $A_2$. The crosses (red in the online version)
  correspond to the four basic states depicted in
  figure~\ref{2Dmovies}.}
\label{critical}
\end{figure}

The bifurcation curves for the different modes in $(St,Re)$-space are
shown in figure~\ref{critical}. At low $St$, mode $B$ is the first to
become critical with increasing $Re$, while at high $St$ mode $A$ is
first. At intermediate values $St \in [40.7,79.7]$, the quasiperiodic
mode bifurcates first, in the form of modulated rotating waves
\MRW. The synchronous mode $A$ always has an azimuthal wave number
$m=2$ ($A_2$), the quasiperiodic mode has $m=1$ ($\MRW_1$), and the
synchronous mode $B$ may have either $m=1$ or $m=2$ depending on
$St$. The bifurcations to the four different states ($B_1$, $B_2$,
$\MRW_1$ and $A_2$) when varying the forcing frequency $St$ are
separated by three codimension-two bifurcation points at which two of
the states bifurcate simultaneously. The four base states shown in
figure~\ref{2Dmovies} correspond to the four distinct bifurcated
states in figure~\ref{critical}.  The synchronous modes for small $St$
have azimuthal wave number $m=1$ ($B_1$) and a single roller fills the
domain most of the time, whereas they have azimuthal wave number $m=2$
($B_2$ and $A_2$) for larger $St>15$ and two rollers persist
throughout the whole forcing cycle. However, the quasiperiodic
$MRW_1$ has azimuthal wave number $m=1$ although it is dominant in a
$St$ region where the two rollers persist.

\begin{figure}
\begin{center}
\includegraphics[width=0.7\linewidth]{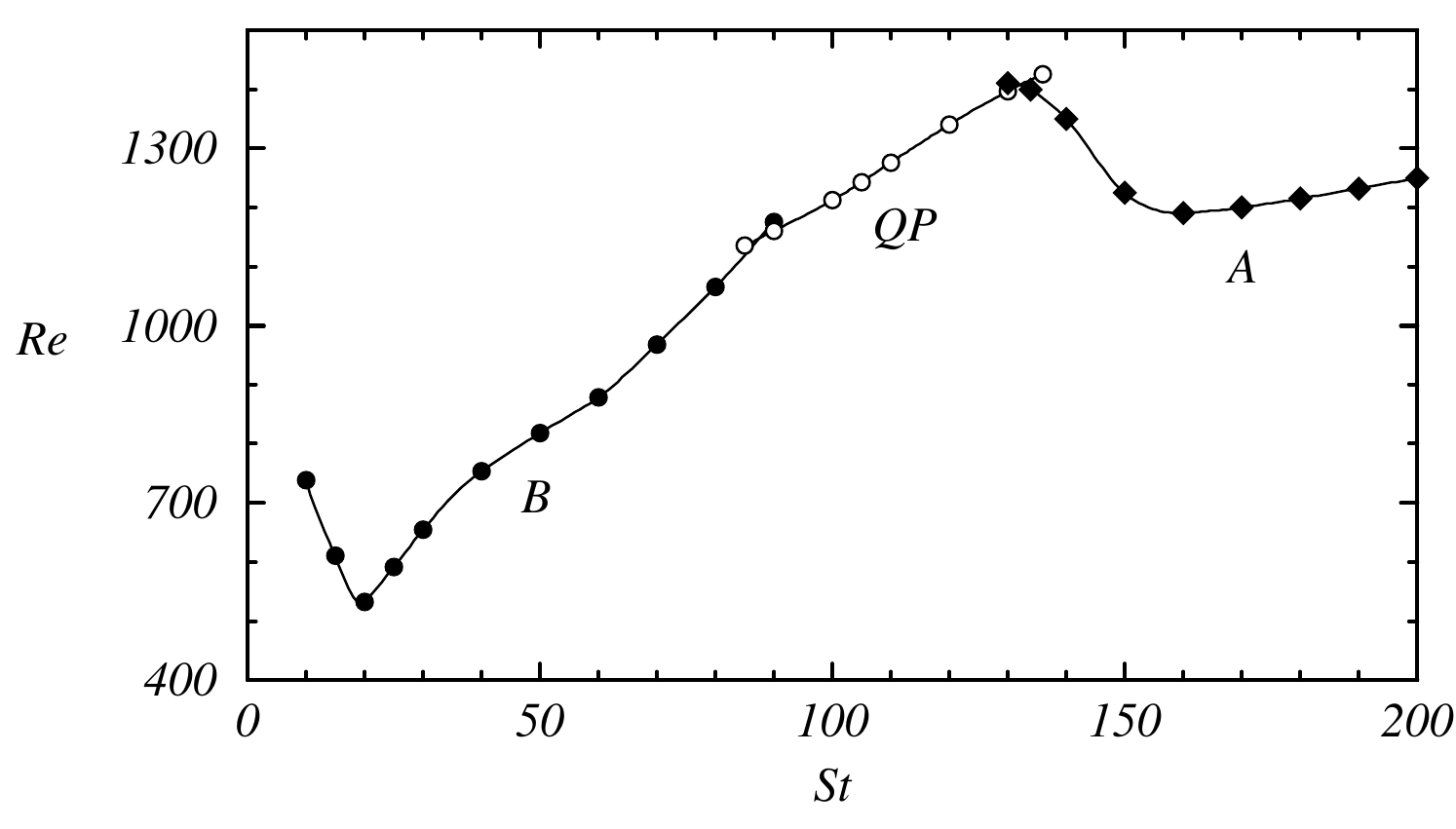}
\end{center}
\caption{Critical Reynolds number, $Re_\textrm{c}$, as a function of
  the Stokes number, $St$, for the transition from the basic state to
  the different three-dimensional states, $B$, $QP$ and $A$, for the
  periodically forced rectangular cavity flow \citep[adapted
    from][]{LHBML05}.}
\label{cavity_Recrit}
\end{figure}

A very similar scenario occurred in the periodically-driven
rectangular cavity problem, as illustrated in
figure~\ref{cavity_Recrit}, showing the critical $Re$ number as a
function of $St$ in the cavity flow \citep[adapted
  from][]{LHBML05}. Figures~\ref{critical} and \ref{cavity_Recrit} are
strikingly similar, and they only differ in their scaling.  The
critical $Re$ and $St$ for the rectangular cavity are about a factor
of two larger than for the cylinder case, so that the marginal curve
in the cavity flow occurs at higher $Re$ number, and the different
modes are shifted to higher $St$. The qualitative shape of the
marginal curves are very similar in both cases, and the shift in
$(St,Re)$ reflects the different geometries of the two problems. An
important difference between the two problems is that in the driven
rectangular cavity the wavenumber of the bifurcated solution varies
continuously, while in the cylinder problem it is discrete (and in
fact of very small wavenumber, either $m=1$ or $m=2$). However, the
qualitative trend is the same in both problems. In the driven
rectangular cavity, the wavenumber of mode $B$ increases with $St$,
while for $QP$ and $A$ their wavenumbers are almost independent of
$St$ \citep{LHBML05}. In the cylinder problem, the azimuthal wave
number of mode $B$ also increases with $St$ (varying from $m=1$ to
$m=2$), while for \MRW\ and $A$ their azimuthal wave numbers do not
vary with $St$ ($m=1$ for \MRW\ and $m=2$ for $A$).

A detailed comparison with the annular cavity problem \citep{BlLo10}
is not possible, because that study focused on the analysis of the
modulated rotating waves, that unfortunately where unstable and
resulted in complicated flows with mixed characteristics between the
synchronous and quasiperiodic solutions. Also, they only considered a
single value of $St=100$ at which the traveling wave state was
expected to be found. Nevertheless, the different modes obtained here
were also present in the annular cavity problem. The radius ratio used
in the annular study was close to one, so that both inner and outer
radii were much larger than the annular gap. That choice was made to
compare with the rectangular cavity flow problem, which corresponds to
the radius-ratio-going-to-one limit. As a result, the azimuthal wave
numbers of the bifurcating states were very large (between $m=30$ and
$m=35$ for the dominant modes in the parameter regime considered). In
contrast, in the cylinder problem which corresponds to radius ratio
zero (the inner cylinder does not exist so that the outer radius and
the gap coincide), the azimuthal wave numbers are very small ($m=1$
and $m=2$). Furthermore, even though the two problems have the same
symmetry group, the flow domain in the cylinder is singly-connected
whereas in the annulus it is doubly-connected.

\section{Three-dimensional structure and symmetries of the unstable
  modes}\label{3Dmodes}

After perturbing the axisymmetric basic flow with
$Re>Re_\textrm{c}(St)$, a new three-dimen\-sional periodic or
quasiperiodic state is reached, after waiting enough time for
saturation. This bifurcated state depends strongly on the mode that
drives the instability, and on the precise values of $(St,Re)$. When
describing these bifurcated flows, we will use the term braid, of
widespread use in similar flows, to denote smaller-scale meridional
structures with vorticity components $\xi$ and $\zeta$. Braids are
typically generated through the amplification of spanwise-orthogonal
perturbations of the rollers in rectangular cavities, and in
cylindrical and annular geometries it is the amplification of
meridional perturbations of the rollers that gives rise to the braids.

In the following subsections the symmetries and features of the
different bifurcated solutions are described and illustrated with
results computed at given values of $St$. Modes $B_1$, $B_2$ and $A_2$
are computed at $St=10$, 32, 100, respectively, whilst mode $\MRW_1$
has been computed at $St=50$, and the corresponding base states have
already been illustrated in figure~\ref{2Dmovies}. All the solutions
have been computed at $Re$ slightly above $Re_\textrm{c}$.

\subsection{Synchronous modes}\label{synch_modes}

\begin{figure}
\begin{center}
\begin{tabular}{cc}
\includegraphics[width=0.48\linewidth]{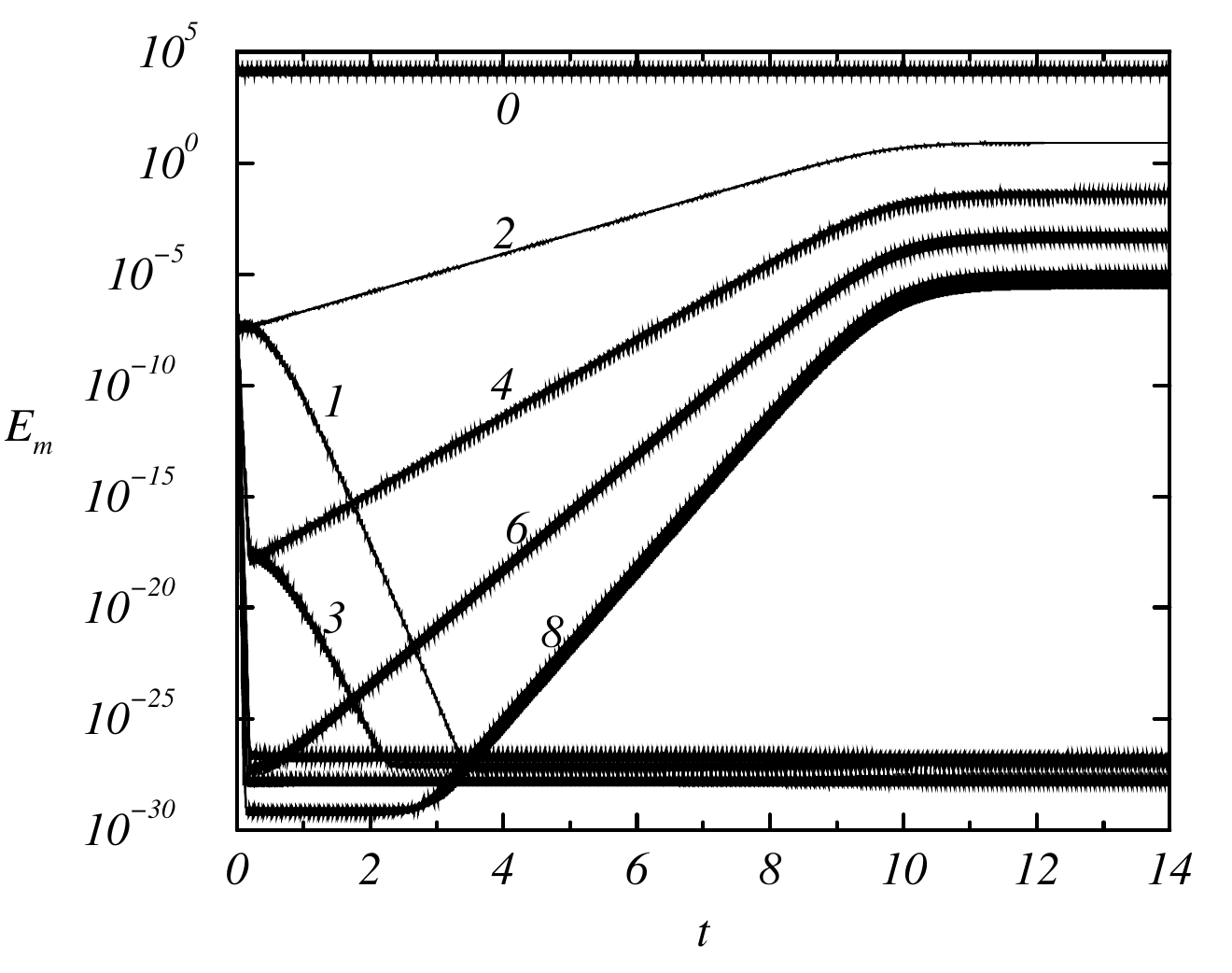} &
\includegraphics[width=0.48\linewidth]{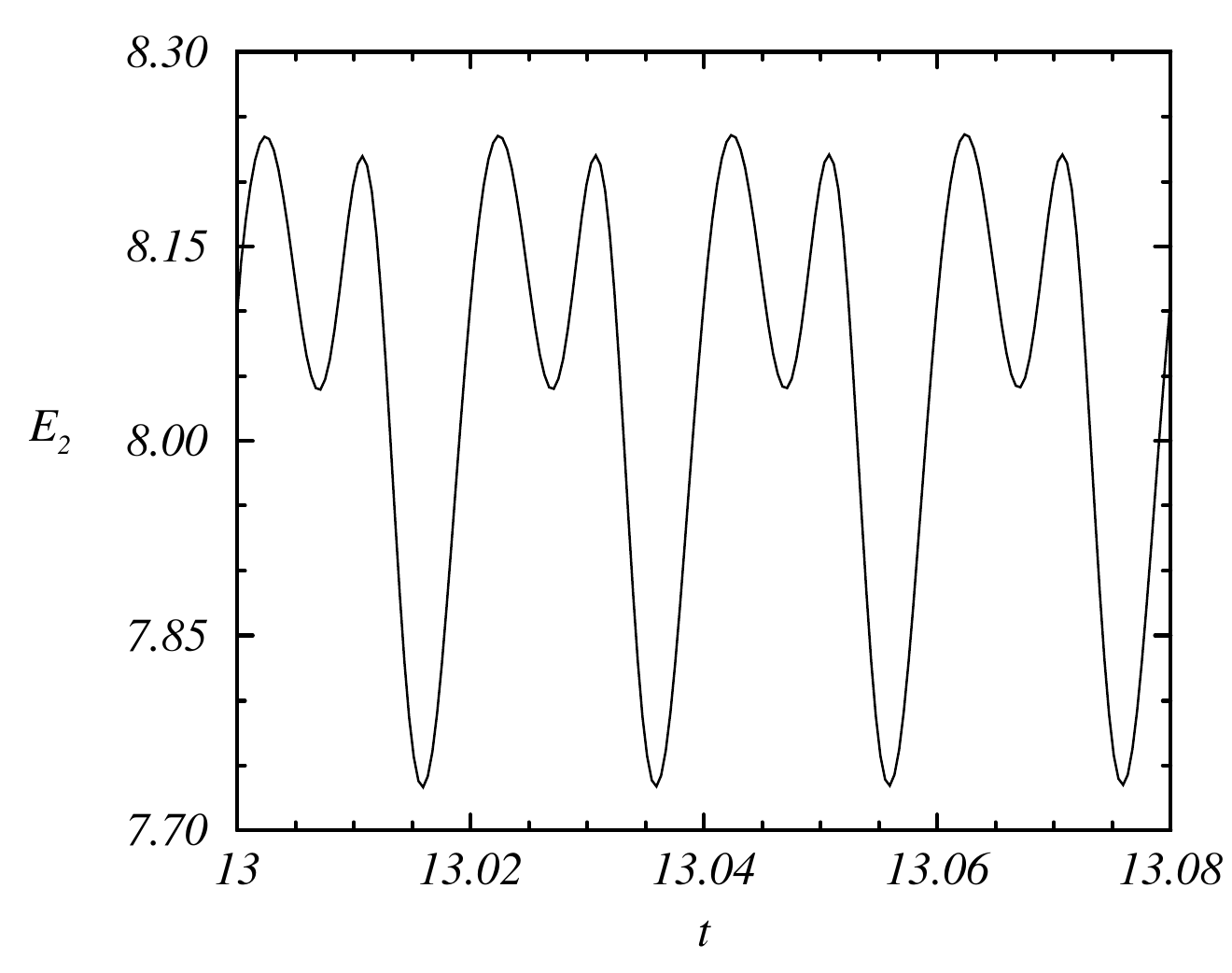}
\end{tabular}
\end{center}
\caption{ $(a)$ Time series of the energies of the leading Fourier
  modes for the $B_2$ state at $Re=525$ and $St=32$; and $(b)$ a
  close-up of $E_2(t)$ after saturation, over two forcing periods.}
\label{time_evol}
\end{figure}

Three-dimensional states result when a single purely real
eigenvalue crosses the unit circle at $+1$ in the complex plane.  When
an axisymmetric flow that belongs to the synchronous region is
perturbed, the energies of the Fourier modes may grow or decay
depending on the case, but what is clear is the modulation of the
energies with the sidewall frequency. When the basic flow is unstable
to synchronous modes, the Fourier spectra begin to grow and at some
time reach an asymptotic state where the modes are saturated but
oscillate with the driving frequency of the wall around a mean
value. Such an evolution can be seen in figure~\ref{time_evol}, where
the energies of the leading Fourier modes are shown as a function of
time for the $B_2$ state at $Re=525$ and $St=32$; the inset show the
oscillations in the energy, synchronous with the forcing (period
$\tau=1/St=0.04$), but with the period halved because the energy is a
sum of squares of the velocities.

\begin{figure}
\begin{center}
\begin{tabular}{ccc}
$(a)$ $B_1$ & $(b)$ $B_2$ & $(c)$ $A_2$ \\
\includegraphics[width=0.32\linewidth]{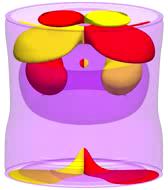} &
\includegraphics[width=0.32\linewidth]{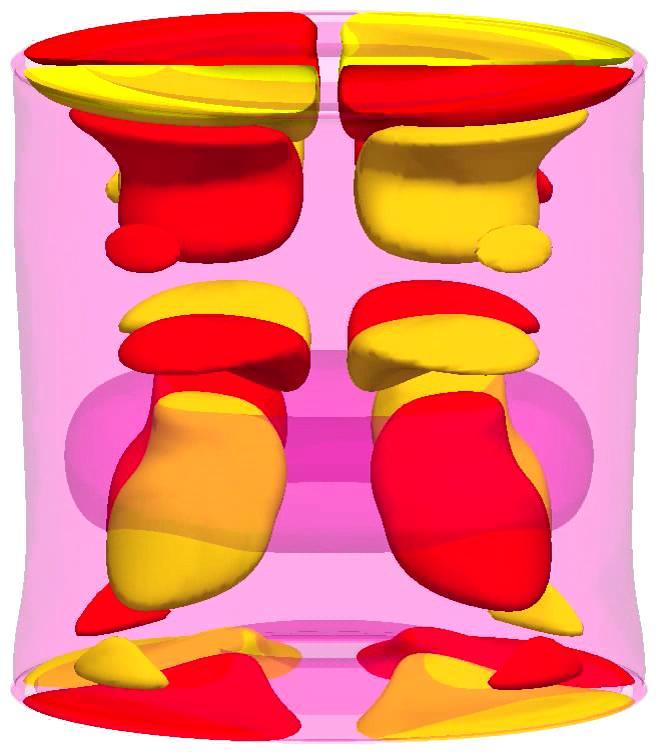} &
\includegraphics[width=0.32\linewidth]{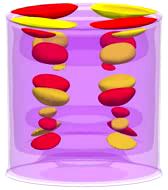}
\end{tabular}
\end{center}
\caption{(Color online) Isosurfaces of radial vorticity $\xi$ (solid)
  and azimuthal vorticity $\eta$ (translucent) for the synchronous
  states $(a)$ $B_1$ at $(Re,St)=(340,10)$ with $\xi=\pm 50$ and
  $\eta=\pm 500$, $(b)$ $B_2$ at $(Re,St)=(525,32)$ with $\xi=\pm 40$
  and $\eta=\pm 1000$, and $(c)$ $A_2$ at $(Re,St)=(700,100)$ with
  $\xi=\pm 150$ and $\eta=\pm 800$.  
  The associated movies online show temporal evolution over one
  forcing period.}
\label{3Dmovies}
\end{figure}

The three-dimensional structures of modes $A$ and $B$ are visualized
in figure~\ref{3Dmovies} with the aid of perspective views of
instantaneous isosurfaces of the radial vorticity (dark/light, or
yellow/red in the online movies, are positive/negative values), which
shows the braid structures, and azimuthal vorticity (translucent),
showing the rollers. Note that the only component of vorticity of the
axisymmetric base state that is non-zero is the azimuthal component,
and that the braids are comprised of radial and axial components of
vorticity and are a direct result of breaking axisymmetry. In general
terms, braids are located near the lids and away from the sidewall,
and they are born on the oblique jets alternatively emerging from the
top and bottom corners. Nevertheless, there are some subtle
variations. For $A_2$, braids suffer slight changes in shape and their
behavior is quite regular as is that of the rollers. Notice that the
shape of each roller stays essentially the same as those of the base
state. For $B_1$ and $B_2$, braids change abruptly during a forcing
cycle, as do the rollers in this regime, and their dynamics (creation,
merging and destruction) are much more complex. In addition, the
azimuthal vorticity of the $B$ modes is very different to that of the
corresponding basic state.

\begin{figure}
\begin{center}
\begin{tabular}{m{10pt}m{0.31\linewidth}m{0.31\linewidth}m{0.31\linewidth}c}
 & \centering $B_1$ & \centering $B_2$ & \centering $A_2$ & \\
$(a)$ & \includegraphics[width=\linewidth]{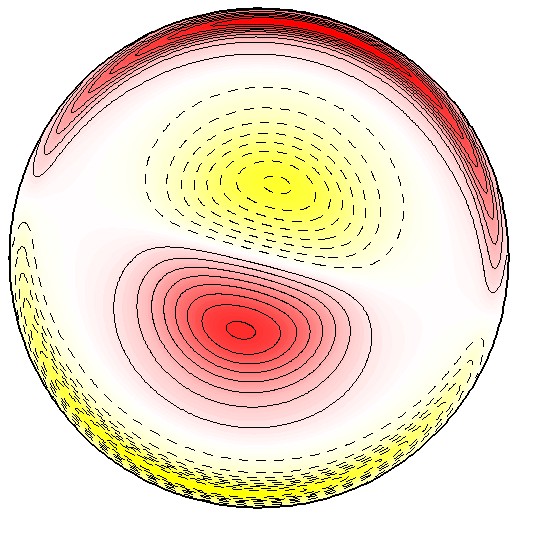}&
\includegraphics[width=\linewidth]{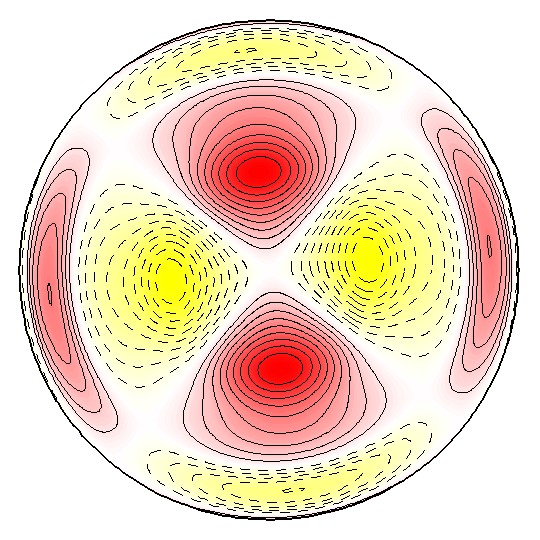}&
\includegraphics[width=\linewidth]{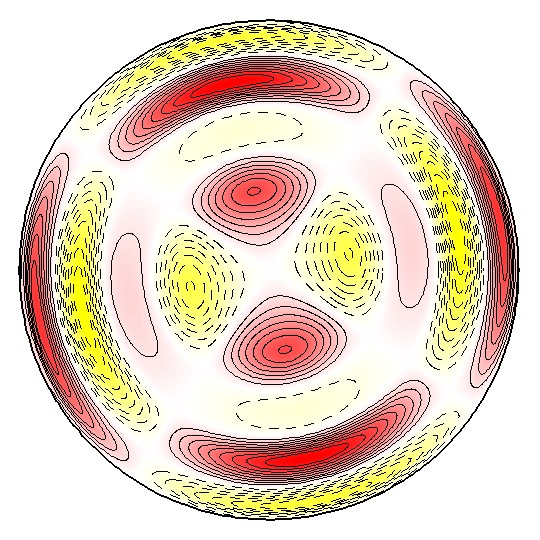}&\\
$(b)$ & \includegraphics[width=\linewidth]{Re340St10_eigenfun_Z001_wz_r_0031}&
\includegraphics[width=\linewidth]{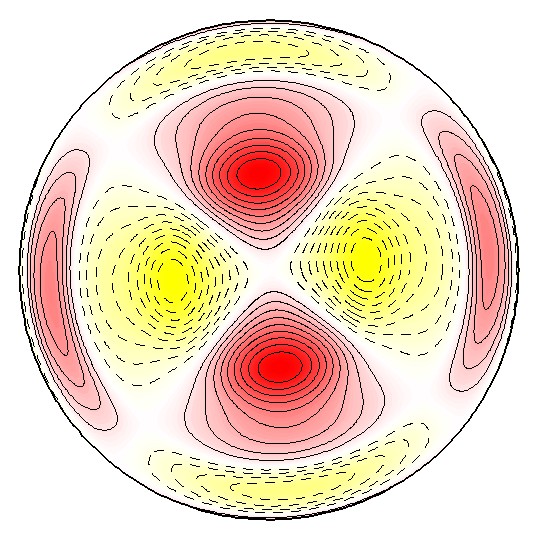}&
\includegraphics[width=\linewidth]{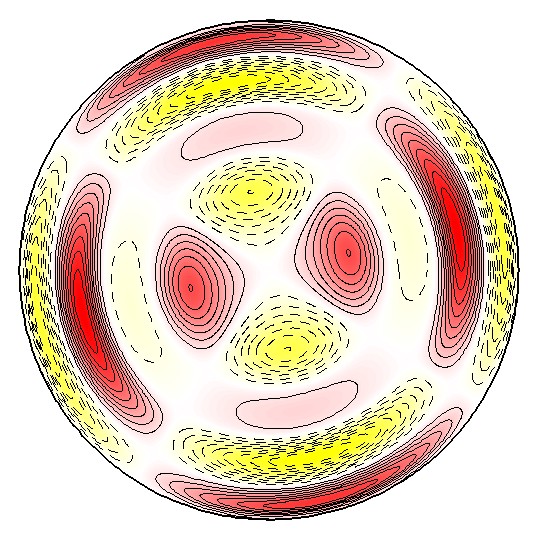}&
\end{tabular}
\end{center}
\caption{(Color online) Axial vorticity contours in a $z$-constant
  plane of the eigenfunctions at the bifurcation to states $B_1$,
  $B_2$ and $A_2$ shown in figure~\ref{3Dmovies}; $(a)$ is computed at
  a convenient time $t_0$ and $z=\varGamma/4$, while $(b)$ is computed
  at time $t_0+\tau/2$ and $z=-\varGamma/4$, i.e.\ applying the
  space-time transformation $H$ to $(a)$. Solid (dashed) contours are
  positive (negative); light/dark (yellow/red online) colors
  correspond to negative/positive values.}
\label{symm_synchr_modes_efun}
\end{figure}

As the bifurcated solutions are no longer axisymmetric, the $O(2)$
symmetry has been broken, and there only remains the discrete symmetry
$R_{2\pi/m}$, a rotation of angle $2\pi/m$ around the axis, where $m$
is the azimuthal wave number of the bifurcated solution. This rotation
generates the so-called $C_m$ (or $Z_m$) symmetry group; when $m=1$
this group is trivial (containing only the identity), and all the
rotational symmetries are destroyed. Now, let us examine what happens
with the spatio-temporal symmetry $H$. We have plotted in
figure~\ref{symm_synchr_modes_efun} axial vorticity contours of the
critical eigenvectors for the $B_1$, $B_2$ and $A_2$ bifurcations in a
horizontal section $z=\varGamma/4$ for a given time, and in the
reflection-symmetric section $z=-\varGamma/4$ after advancing half the
forcing period. The figure shows that the bifurcations to $B_1$ and $B_2$ are
$H$-symmetric, i.e.\ the values of the axial vorticity $\zeta$ of the
eigenfunctions, at a
given time $t_0$ and at $z=\varGamma/2$
(figure~\ref{symm_synchr_modes_efun}$a$), are the same as the values
of $\zeta$ advancing time by half the forcing period, $t_0+\tau/2$, on
the reflection-symmetric plane $z=-\varGamma/2$
(figure~\ref{symm_synchr_modes_efun}$b$). The eigenfunction of the
$A_2$ bifurcation is not $H$-symmetric, but changes sign, so the
$H$-symmetry is broken in this bifurcation. However, $H$ combined with
the rotation $R_{\pi/m}$, with $m=2$ (half the angle of the rotational symmetry
of the state), results in a space-time symmetry of the
$A_2$ eigenfunction. This is precisely the expected behavior from
bifurcation theory \citep{MLB04}: there are only two options for
three-dimensional synchronous eigenfunctions under the action of the
space-time symmetry $H$, multiplication by $\pm1$. The behavior of all
the velocity and vorticity components is given by
\begin{align}
&\text{$H$ preserved: }&&\left\{\begin{array}{l}
 (u,v,w)(r,\theta,z,t)=(u,v,-w)(r,\theta,-z,t+\tau/2), \\
 (\xi,\eta,\zeta)(r,\theta,z,t)=(-\xi,-\eta,\zeta )(r,\theta,-z,t+\tau/2),
 \end{array}\right. \label{preserved_H_sym} \\
&\text{$H$ broken: }&&\left\{\begin{array}{l} (u,v,w)_\textrm{e}(r,\theta,z,t)=
 (-u,-v,w)_\textrm{e}(r,\theta,-z,t+\tau/2), \\
 (\xi,\eta,\zeta)_\textrm{e}(r,\theta,z,t)=(\xi,\eta,-\zeta)_\textrm{e}
 (r,\theta,-z,t+\tau/2), \end{array}\right. \label{broken_H_sym} \\
&\text{$R_{\pi/m}H$ preserved:}&&\left\{\begin{array}{l}
 (u,v,w)(r,\theta,z,t)=(u,v,-w)(r,\theta+\pi/m,-z,t+\tau/2), \\
 (\xi,\eta,\zeta)(r,\theta,z,t)=(-\xi,-\eta,\zeta)(r,\theta+\pi/m,-z,t+\tau/2),
 \end{array}\right. \label{preserved_RH_sym}
\end{align}

\begin{figure}
\begin{center}
\begin{tabular}{m{10pt}m{0.31\linewidth}m{0.31\linewidth}m{0.31\linewidth}c}
 & \centering $B_1$ & \centering $B_2$ & \centering $A_2$ & \\
$(a)$ & \includegraphics[width=\linewidth]{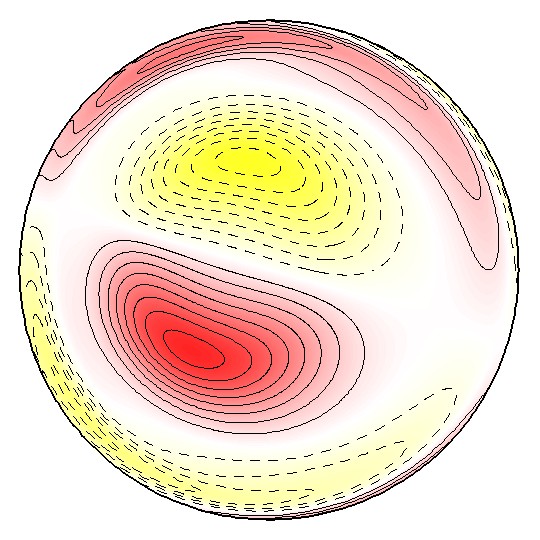}&
\includegraphics[width=\linewidth]{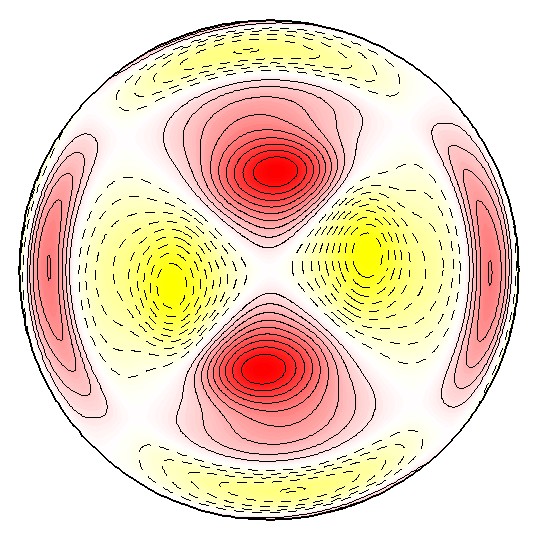}&
\includegraphics[width=\linewidth]{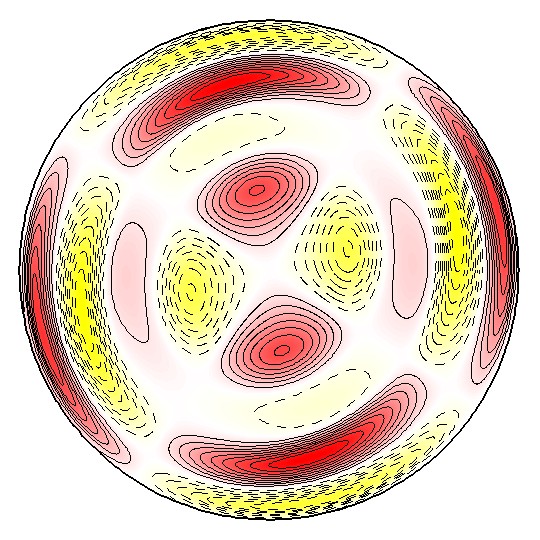}&\\
$(b)$ & \includegraphics[width=\linewidth]{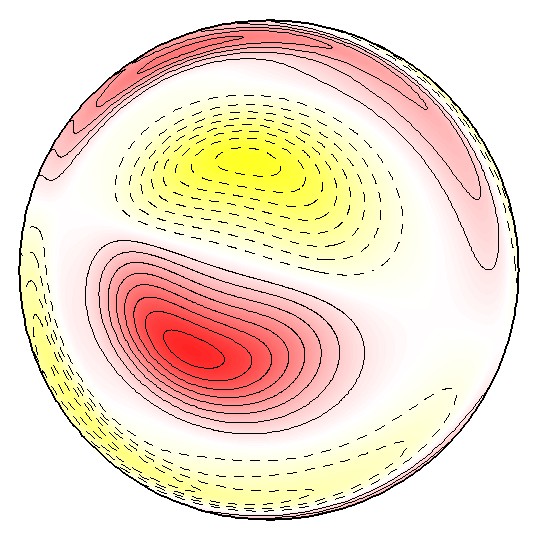}&
\includegraphics[width=\linewidth]{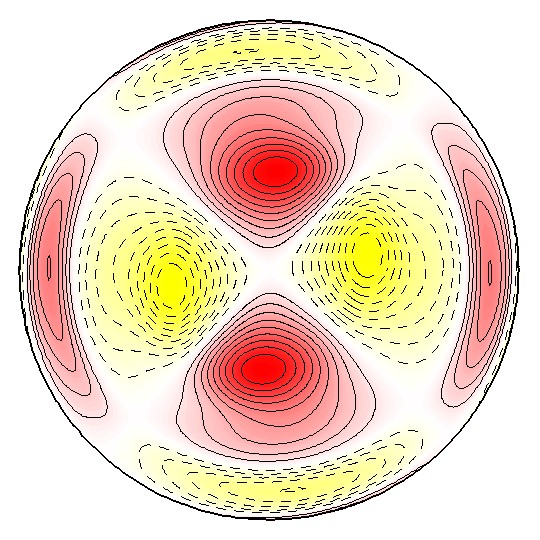}&
\includegraphics[width=\linewidth]{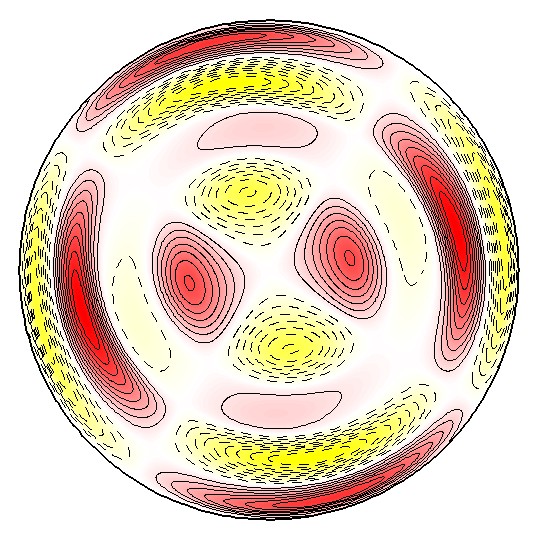}&
\end{tabular}
\end{center}
\caption{(Color online) Axial vorticity contours in a $z$-constant
  plane of nonlinear saturated states corresponding to $B_1$, $B_2$
  and $A_2$ shown in figure~\ref{3Dmovies}; $(a)$ is computed at a
  convenient time $t_0$ and $z=\varGamma/4$, while $(b)$ is computed
  at time $t_0+\tau/2$ and $z=-\varGamma/4$, i.e.\ applying the
  space-time transformation $H$ to $(a)$. Solid (dashed) contours are
  positive (negative); light/dark (yellow/red online) colors
  correspond to negative/positive values.}
\label{symm_synchr_modes}
\end{figure}

The space-time symmetries of the eigenfunctions translates to the
nonlinear saturated states (as long as no additional bifurcations take
place in the saturation process). However, the multiplication by -1
shown in \eqref{broken_H_sym} is a property of the eigenfunction that
the saturated states do not have. The reason is that the
eigenfunctions are pure Fourier modes in the azimuthal direction, and
when they develop to fully nonlinear three-dimensional bifurcated
solutions, Fourier harmonics appear, and the even harmonics (including
the zero mode) are multiplied by $(-1)^2=+1$ under the action of the
$H$-symmetry, so the full nonlinear solution does not have the
multiplication by -1 property that the eigenfunction has.
Figure~\ref{symm_synchr_modes} shows the same information as in
figure~\ref{symm_synchr_modes_efun}, but for the full nonlinear
bifurcated solutions, illustrating the symmetry properties of the
saturated states. The behavior labeled by the subscript $\textrm{e}$
in \eqref{broken_H_sym}, corresponding to multiplication by -1 under
the action of the $H$-symmetry, is no longer present in the saturated
nonlinear solution. However, the preserved symmetries $H$
\eqref{preserved_H_sym} and $R_{\pi/m}H$ \eqref{preserved_RH_sym}
clearly persist.

\begin{figure}
\begin{center}
\begin{tabular}{ccc}
$(a)$ B1 & $(b)$ B2 & $(c)$ A2 \\ 
\includegraphics[width=0.32\linewidth]{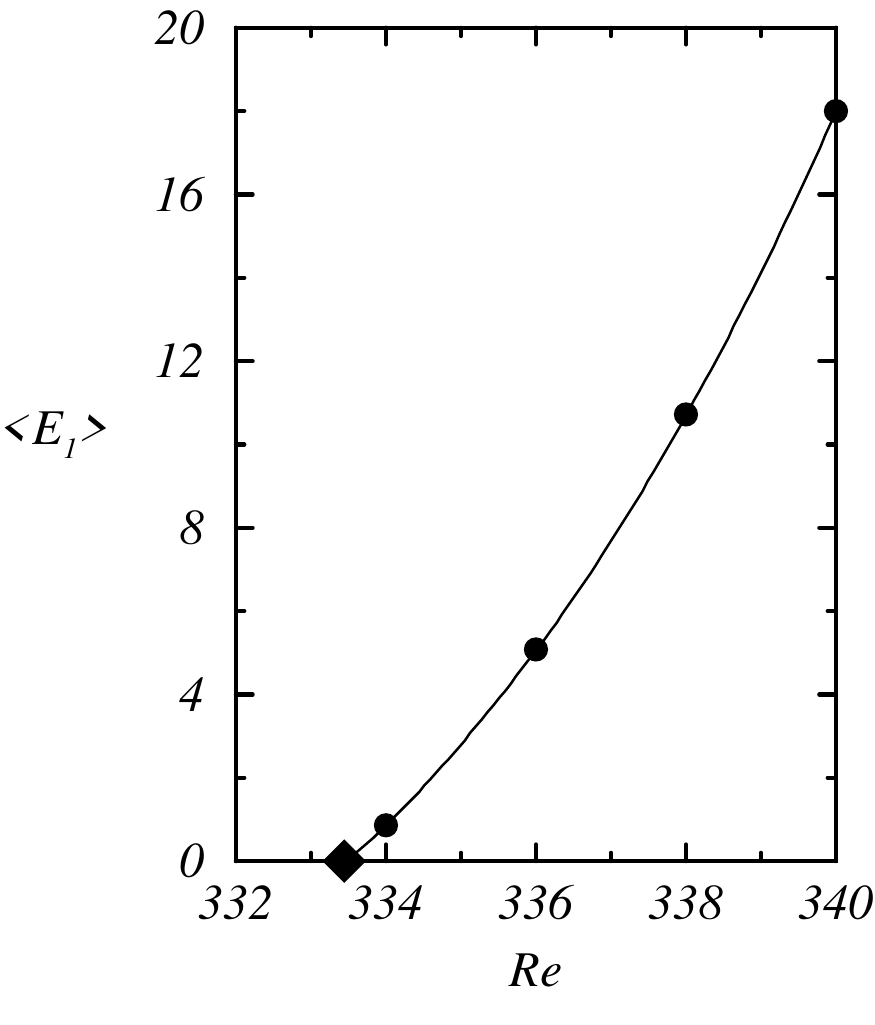} &
\includegraphics[width=0.32\linewidth]{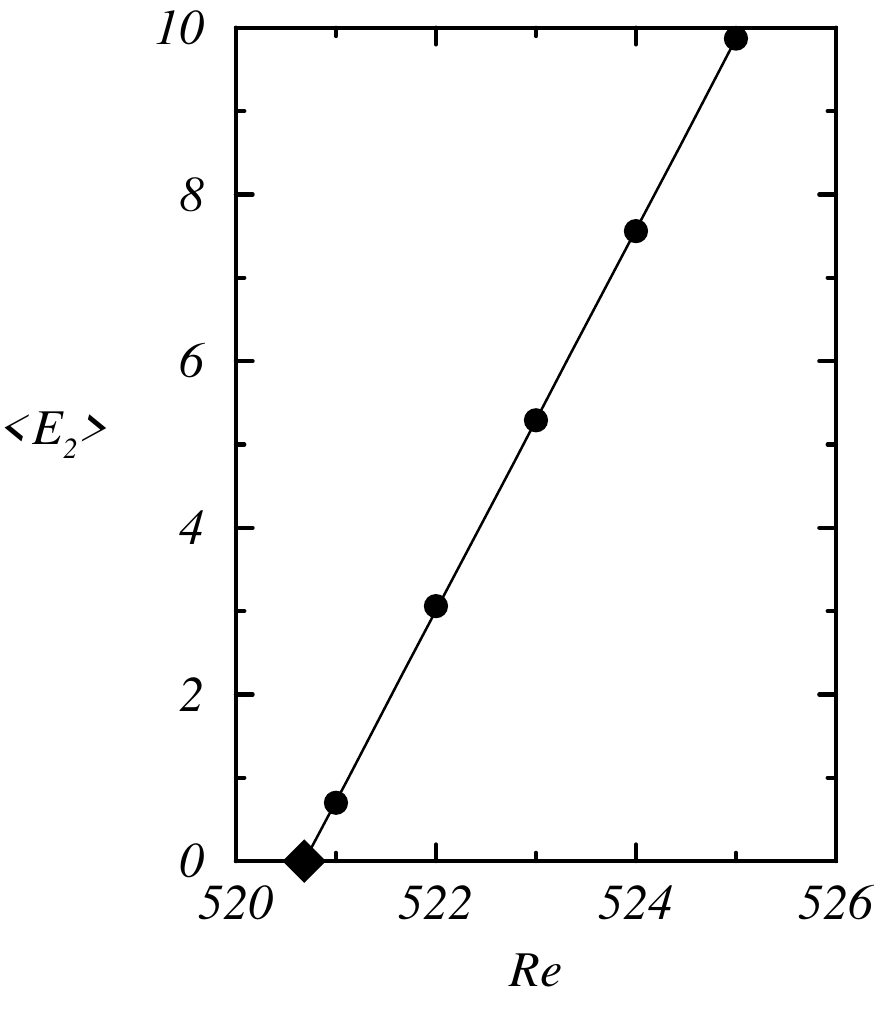} &
\includegraphics[width=0.32\linewidth]{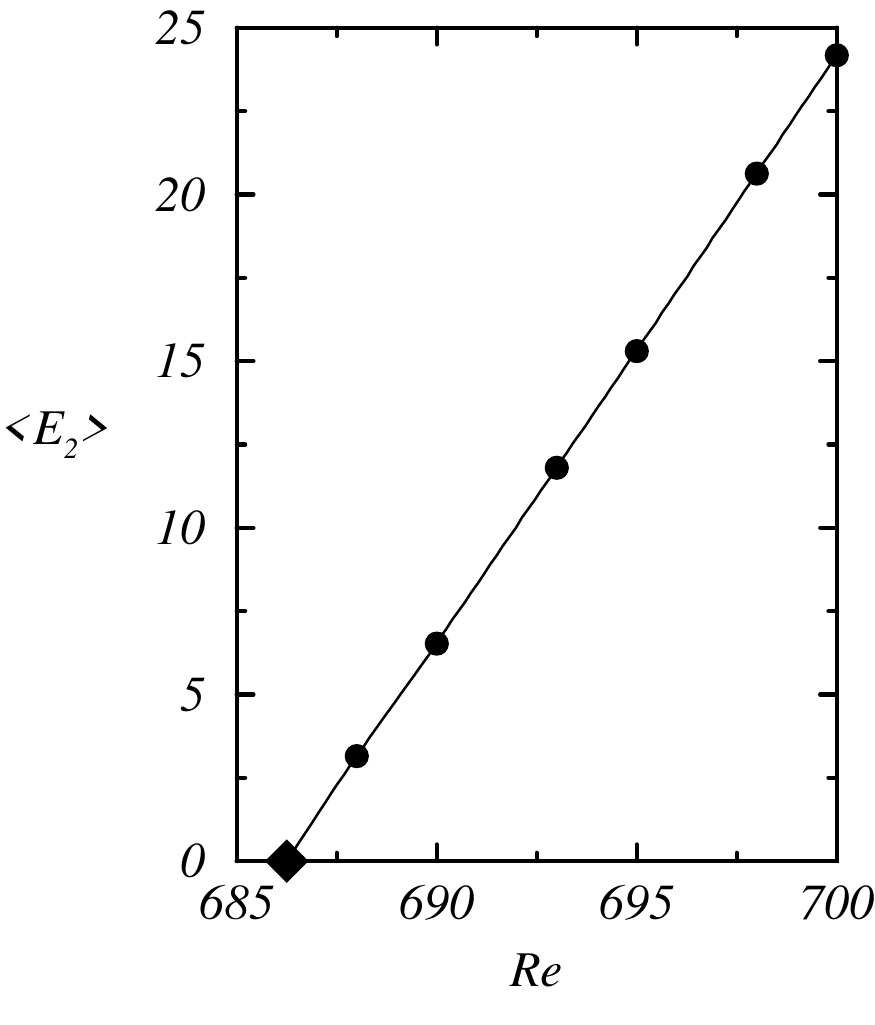}
\end{tabular}
\end{center}
\caption{Time average of the energy of the dominant mode for the
  synchronous solutions as a function of $Re$, $(a)$ B1 solutions at
  $St=10$, $(b)$ B2 solutions at $St=$, and $(c)$ A2 solutions at
  $St=100$. The diamond corresponds to the critical $Re$ obtained by
  linear stability analysis and the solid disks correspond to computed
  nonlinear solutions.}
\label{supercrit_synchronous}
\end{figure}

The three bifurcations to $B_1$, $B_2$ and $A_2$ are supercritical, as
shown in figure~\ref{supercrit_synchronous}, where the time-averaged
energy of the dominant mode $\langle
E_m\rangle$ is plotted as a function of the Reynolds
number; there is no hysteresis, and the behavior of $\langle
E_m\rangle$ is linear as $Re$ approaches the critical value
($\blacklozenge$). The normal form for the amplitude of the bifurcated
synchronous solutions in the supercritical case is given by $\dot
A=A(\mu-c|A|^2)$. When saturation is reached, $\dot A=0$ and we have
$Re-Re_\textrm{c}=\mu=c|A|^2=d\langle E_m\rangle$, the observed
linear behavior close to the bifurcation point.

\subsection{Quasiperiodic mode}\label{QPmodes}

The onset of the quasiperiodic states occurs when two
complex-conjugate pairs of eigenvalues cross the unit circle, thus
introducing a second frequency $\omega_Q$ related to the phase of the
complex-conjugate pairs. Generically this second frequency will be
incommensurate with the forcing frequency, so the $H$-symmetry is
broken in this bifurcation. The second frequency $\omega_Q$ can
manifest itself in two ways, depending on whether the bifurcation
breaks $K_\theta$ or $R_\alpha$ of the $O(2)$ symmetry of the basic
state. In the linear stability analysis we referred only to mode
$\MRW_1$, but this term encompasses modulated $\theta$-travelling wave
(\MRW) and modulated standing wave (\MSW) states. Due to the $O(2)$
symmetry of the governing equations, there are two pairs of complex
conjugate eigenvalues that bifurcate simultaneously, and they
correspond to modulated $\theta$-travelling waves, that can travel in
the positive or negative $\theta$-direction; after a period of the
forcing, the flow pattern repeats itself, but rotated a certain angle,
$\pm\theta_0$, related to the second frequency by
$\theta_0=2\pi\omega_Q/St$, where $St=\omega_f$ is the forcing
frequency. $K_\theta$ transforms each one of the \MRW\ into the
other, therefore the $K_\theta$-symmetry is broken; the $SO(2)$
rotational symmetry is also broken, because the solution has azimuthal
wave number $m=1$ ; however, as they are modulated traveling waves,
there is a preserved space-time symmetry, consisting in advancing one
forcing period in time combined with the rotation $R_{\pm\theta_0}$.

Besides the two \MRW\ solutions, there is also a third nonlinear
solution corresponding to a symmetric combination of the two
\MRW\ states; these states, called modulated standing waves \MSW,
are $K_\theta$-symmetric, but the $SO(2)$ rotational symmetry is
completely broken. Only one of the two families of solutions, \MRW and
\MSW, is stable \citep{MLB04}. In the present problem, as in the
case for the driven annular cavity \citep{BlLo03_jfm} and for the
driven annular cavity \citep{LHBML05}, the stable solutions are \MRW;
their sense of travel depends on the initial condition for the sign of
the azimuthal velocity perturbation. In order to obtain \MSW, it is
necessary to enforce the $K_\theta$ symmetry, restricting the
computations to the appropriate Fourier subspace.

When an axisymmetric flow unstable to $\MRW_1$ mode is perturbed, the
energies of all Fourier modes begin to grow with the driving frequency
and this is additionally modulated by the quasiperiodic frequency,
which is approximately one order of magnitude smaller than the forcing
frequency, and it is the same for \MRW\ and \MSW\ states. However,
when \MSW\ reaches a saturated state, the energy of the
Fourier modes retain both characteristic times, whilst the
quasiperiodic frequency for \MRW, being related to the azimuthal
precession of the pattern, does not manifest in the energy of the
Fourier modes.

\begin{figure}
\begin{center}
\begin{tabular}{c@{\hspace{20pt}}c}
$(a)$ $\MRW_1$ & $(b)$ $\MSW_1$ \\
\includegraphics[width=0.33\linewidth]{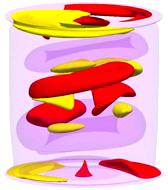} &
\includegraphics[width=0.33\linewidth]{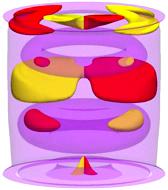}
\end{tabular}
\end{center}
\caption{(Color online) Isosurfaces of radial vorticity $\xi$ (solid)
  and azimuthal vorticity $\eta$ (translucent) for $(a)$ $\MRW_1$ with
  $\xi=\pm 180$, and $(b)$ $\MSW_1$ with $\xi=\pm 120$, both with
  $\eta=\pm 1000$ and computed at $(Re,St)=(615,50)$; only $\MRW_1$ is
  stable. Light/dark (yellow/red online) solid colors correspond to
  negative/positive values of $\xi$. The associated movies online show
  temporal evolution over several forcing periods.}
\label{3D_QPmovies}
\end{figure}

The three-dimensional structures of these quasiperiodic flows are
visualized in figure~\ref{3D_QPmovies} by means of perspective views
of instantaneous isosurfaces of the axial vorticity (dark/light are
positive/negative values) and azimuthal vorticity
(translucent). Braids are concentrated on the cylinder endwalls away
from the sidewall and suffer large variations in all cases. As with
the synchronous modes, the braids seem to be born along the oblique
jets emerging from the corners and propagate into the interior,
interacting in a complex way with the braids coming from the other
endwall. Nevertheless, for \MSW\ the braids possess very regular
shapes and look quite similar to those of the synchronous states, and
the rollers are virtually not distorted. For \MRW\ however, the braids
have a helical structure and the rollers do not resemble the
corresponding base flow rollers at all. In fact, the rollers in
\MRW\ are tilted with respect to the horizontal rollers in the base
state.

\begin{figure}
\begin{center}\setlength{\pacolength}{0.24\textwidth}
\begin{tabular}{m{5pt}m{0.22\textwidth}m{0.22\textwidth}m{0.22\textwidth}
   m{0.22\textwidth}c}
 & \centering $t=0$ & \centering $t=10T$ & \centering $t=20T$ &
   \centering $t=30T$ & \\
\rotatebox{90}{$\MRW_1$} & 
\includegraphics[width=\pacolength]{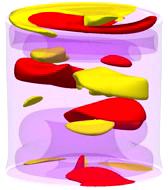} &
\includegraphics[width=\pacolength]{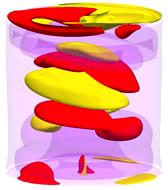} &
\includegraphics[width=\pacolength]{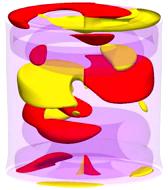} &
\includegraphics[width=\pacolength]{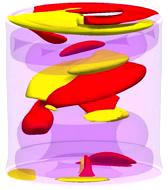} & \\
\rotatebox{90}{$\MSW_1$} & 
\includegraphics[width=\pacolength]{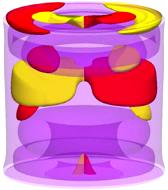} &
\includegraphics[width=\pacolength]{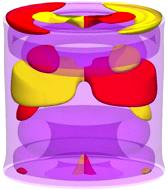} &
\includegraphics[width=\pacolength]{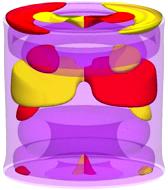} &
\includegraphics[width=\pacolength]{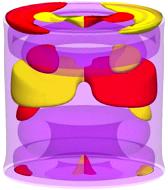} &
\end{tabular}
\end{center}
\caption{(Color online) Same solutions and isosurface levels as in
  figure~\ref{3D_QPmovies}, but strobed every 10 forcing periods. The
  associated movies online show temporal evolution strobed every
  forcing period.}
\label{TWstrobed}
\end{figure}

Figure~\ref{TWstrobed} shows the same contours as in
figure~\ref{3D_QPmovies}, all at the same phase at integer multiples
of the forcing period apart. For \MRW, the strobed structures do not
change in a frame of reference that rotates in the azimuthal by an
angle $\theta_0$ every forcing period, justifying the name of
modulated rotating wave. For the \MRW\ shown in
figures~\ref{3D_QPmovies} and \ref{TWstrobed} at $(Re,St)=(615,50)$,
this value is $\theta_0\approx28.77^\circ$. This results in a
precession frequency $\omega_p=3.996$. For \MSW, the strobed
structures vary substantially in one period, as can be seen in the
movie associated with figure~\ref{TWstrobed}$(b)$. In general, the
ratio between the quasiperiodic and wall periods are not commensurate,
and the flow structure never repeats itself. However, the flow
structure of \MSW\ remains almost unchanged after ten forcing periods,
as shown in figure~\ref{TWstrobed}$(b)$; the only noticeable
difference is the formation of braids very close to the bottom
lid. This is because the ratio of quasiperiodic to forcing frequencies
is very close to $1/10$ for the parameter values $(Re,St)=(615,50)$ of
\MSW; this will be explored in more detail at the end of the present
section.

\begin{figure}
\begin{center}
\includegraphics[width=0.7\linewidth]{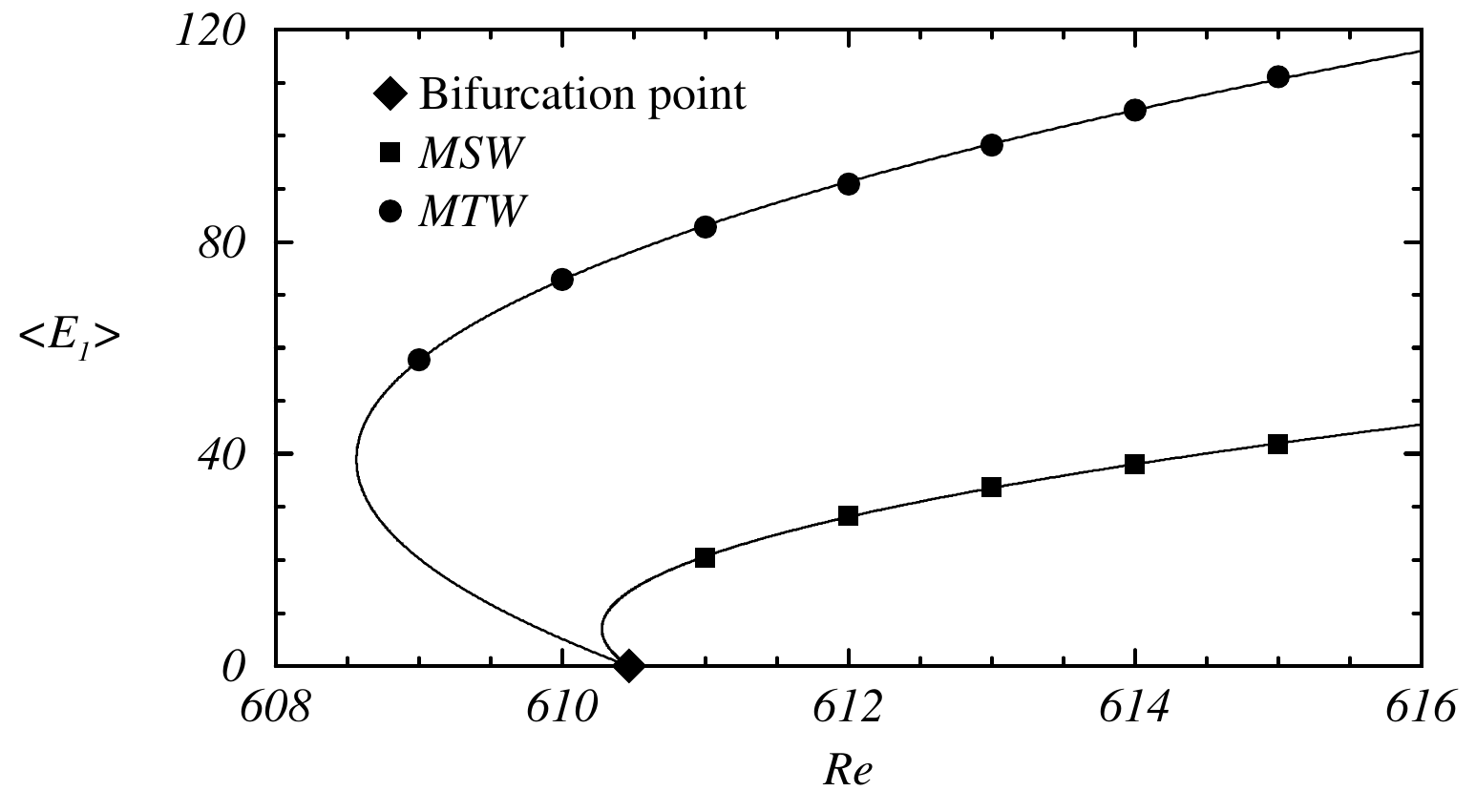}
\end{center}
\caption{Time average of the energy of the $m=1$ dominant mode for the
  quasiperiodic solutions as a function of $Re$, for $St=50$. The
  diamond corresponds to the critical $Re$ obtained by linear
  stability analysis. Symbols correspond to computed saturated
  solutions, and the curves are best fits to the parabolic profiles
  predicted by normal form theory.}
\label{subcrit_QP}
\end{figure}

The quasiperiodic bifurcation is subcritical, for both \MRW\ and \MSW,
in contrast to the synchronous bifurcations which are
supercritical. Figure~\ref{subcrit_QP} shows the time average of the
energy of the $m=1$ dominant mode for the quasiperiodic solutions,
$\langle E_1\rangle$, as a function of the Reynolds number. The
\MRW\ solutions show a well-defined hysteretic region; for the \MSW,
the hysteretic region is very costly computationally to obtain,
having extremely long transients. Nevertheless, the energies do not
behave linearly close to the critical Reynolds $Re_\textrm{c}=610.47$
for $St=50$. We can fit the computed energy amplitudes to the shape
predicted by normal form theory. According to \citet{MLB04}, the
amplitudes of the bifurcated solutions vary as
\begin{equation}
 Re-Re_\textrm{c}=\mu=a|A|^4-b|A|^2=a\langle E_1\rangle^2-b\langle
 E_1\rangle,
\end{equation}
where a quartic term has been included due to the subcritical nature
of the bifurcation. The fitting parameters $a$ and $b$ can be
expressed in terms of the energy and $Re$ at the saddle-node point:
\begin{equation}
  \frac{Re-Re_\textrm{c}}{Re_\textrm{c}-Re_\mathit{SN}}=
  \frac{\langle E_1\rangle}{\langle E_1\rangle_\mathit{SN}}\biggr(
  \frac{\langle E_1\rangle}{\langle E_1\rangle_\mathit{SN}}-2\biggr).
\end{equation}
The solid lines in figure~\ref{subcrit_QP} are best fits of this
expression to the computed values. The agreement is very good and
provides good estimates of the Reynolds numbers of the saddle-node
bifurcations. The estimates are $Re_\mathit{SN}\approx610.28$ for
\MSW\ and $Re_\mathit{SN}\approx608.57$ for \MRW.

\begin{figure}
\begin{center}
\begin{tabular}{cc}
 $(a)$ & $(b)$ \\
 \includegraphics[width=0.48\linewidth]{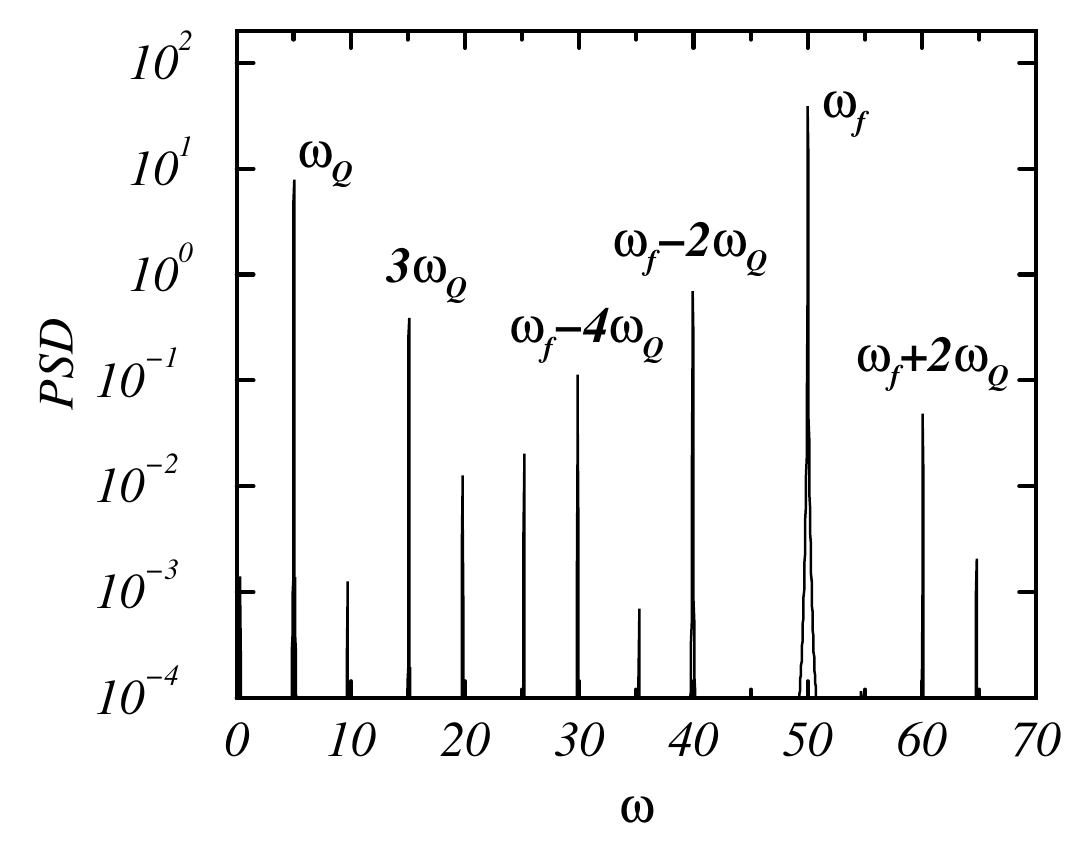} & 
 \includegraphics[width=0.48\linewidth]{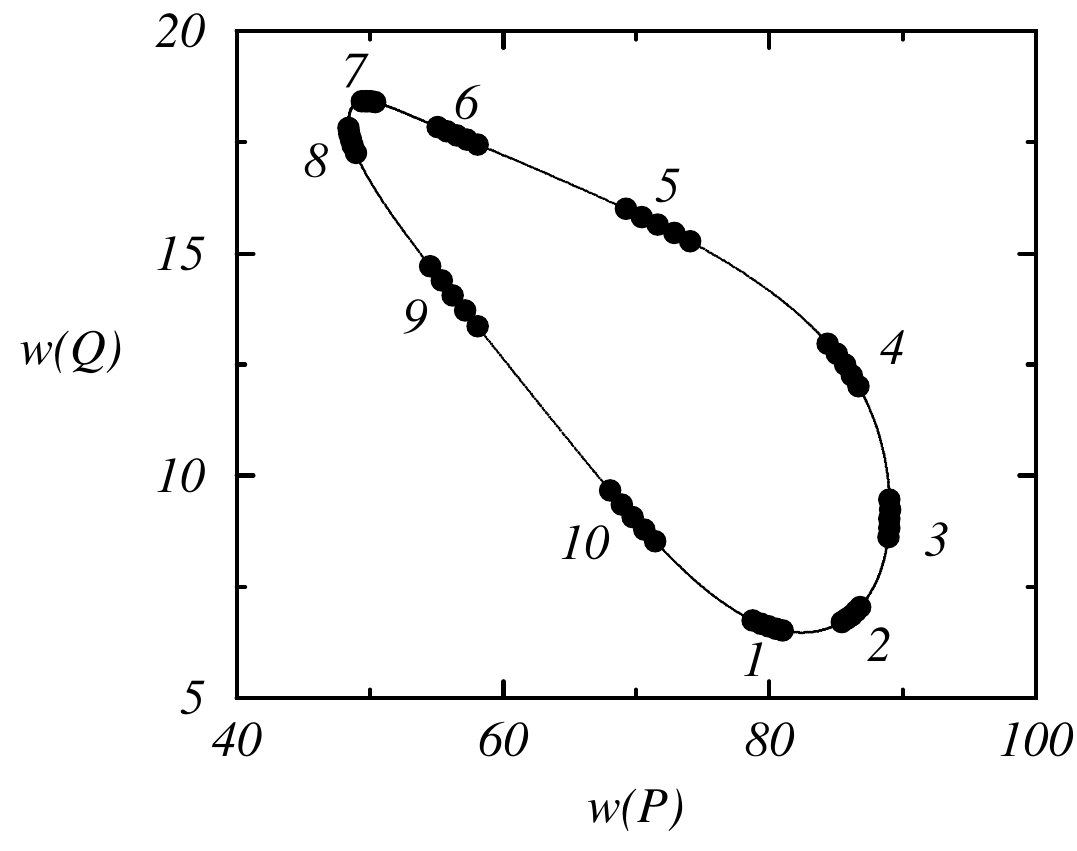}
\end{tabular}
\end{center}
\caption{$(a)$ Power spectral density of the FFT for \MSW\ at
  $(Re,St)=(615,50)$ and $(b)$ Poincar\'e section of the same solution
 strobing \MSW\ every $\tau$. Symbols {\large$\bullet$} and numbers in
 $(b)$ corresponds to successive iterates, showing that the frequency
 ratio is close to rational: $\omega_Q/\omega_f\approx1/10$.}
\label{Poincare_FFT_MSW}
\end{figure}

The frequencies of the quasiperiodic states can be computed via FFT of
the time series of a convenient variable; here we have chosen the
value of the axial velocity at a point $P$ close to the sidewall at
the cylinder mid-height,
$(r,\theta,z)_P=(0.9,0,0)$. Figure~\ref{Poincare_FFT_MSW}$(a)$ shows
the power spectral density for \MSW\ at $(Re,St)=(615,50)$. It is a
quasiperiodic spectrum with two well-defined frequencies,
$\omega_f=50$ and $\omega_Q=5.035$, and their linear combinations. The
ratio of the frequencies is close to resonance,
$\omega_Q/\omega_f=0.1007\approx1/10$, and so a small frequency
corresponding to $10\omega_Q-\omega_f\approx0.353$ is also present.
The FFT supplies $\omega_Q$ up to a multiple of the forcing frequency;
its precise value must be obtained by other methods. In order to
analyze in detail how close to resonance the Neimark--Sacker
bifurcation is and also to confirm the value of the second frequency,
we have computed Poincar\'e sections of the quasiperiodic states by
strobing the \MSW\ solution every period $\tau$. The Poincar\'e
section is shown in Figure~\ref{Poincare_FFT_MSW}$(b)$, where we have
projected the infinite-dimensional phase space into the plane
corresponding to the values of the axial velocity $w$ at two different
locations $P$ and $Q$ in the cylindrical domain. The point $P$ is the
same used for computing the FFT, at $(r,\theta,z)_P=(0.9,0,0)$, and
$Q$ is close to the bottom endwall at
$(r,\theta,z)_P=(0.9,0,-0.1)$. The closed curve is the section of the
two-torus where the solution lives, and the symbols {\large$\bullet$}
correspond to successive iterates that are numbered in the figure. The
tenth iterate, after $10\tau$, almost coincides with the initial
point. We have plotted 50 iterates, so we have 10 clusters of 5
points, showing that the rotation number (the ratio of the frequencies
$\omega_Q/\omega_f$) is very close to $1/10$. This justifies the
selection of $\omega_Q=5.035$ from the FFT.

\begin{figure}
\begin{center}
\begin{tabular}{cc}
 \includegraphics[width=0.48\linewidth]{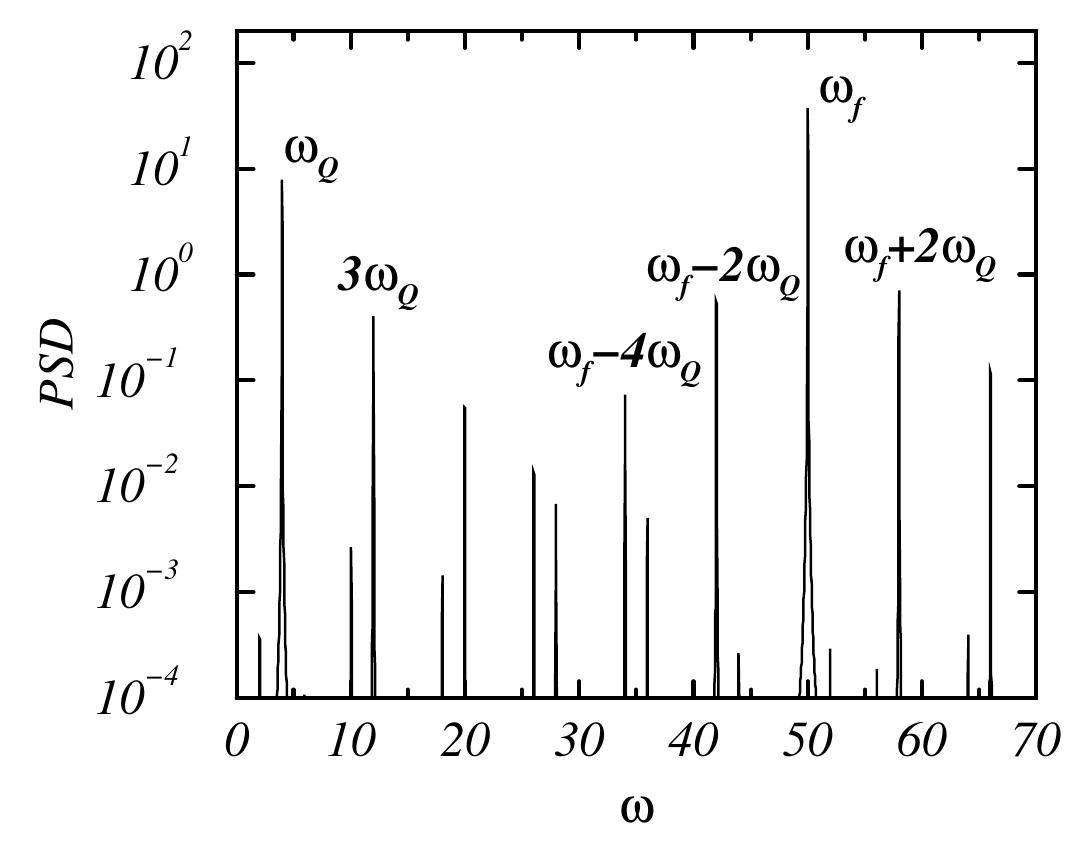} & 
 \includegraphics[width=0.48\linewidth]{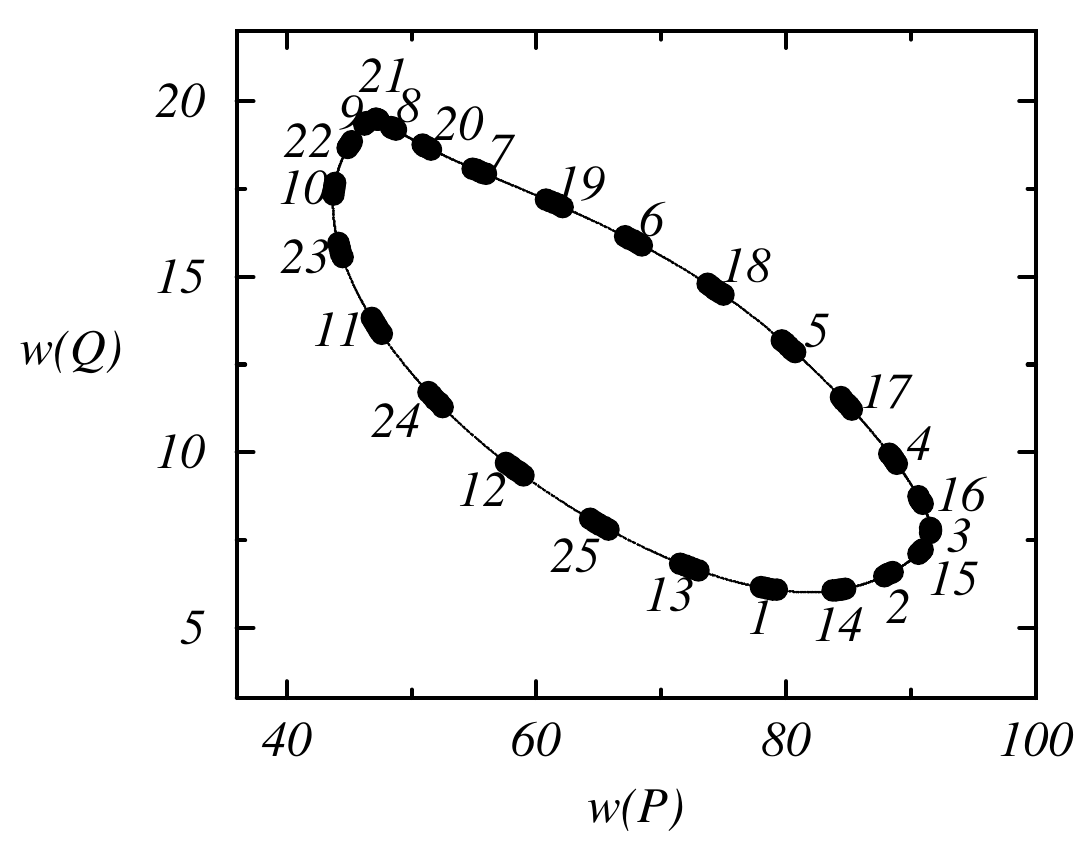}
\end{tabular}
\end{center}
\caption{$(a)$ Power spectral density of the FFT for \MRW\ at
  $(Re,St)=(615,50)$ and $(b)$ Poincar\'e section of the same solution
 strobing \MRW\ every $\tau$. Symbols {\large$\bullet$} and numbers in
 $(b)$ corresponds to successive iterates, showing that the frequency
 ratio is close to rational: $\omega_Q/\omega_f\approx 2/25$.}
\label{Poincare_FFT_MRW}
\end{figure}

Figure~\ref{Poincare_FFT_MRW}$(a)$ shows the power spectral density
for \MRW\ at $(Re,St)=(615,50)$. The second frequency is
$\omega_Q=3.996$, quite different and smaller than the frequency of
\MSW.  In this case we are even closer to resonance, but a different
one. From the Poincar\' e section in
figure~\ref{Poincare_FFT_MSW}$(b)$, we see that the iterates undergo
two turns on the section before almost coinciding with the initial
point after 25 iterates, so now the rotation number (frequency ratio)
is $\omega_Q/\omega_f=0.07992\approx2/25$, and a very small frequency
appears, $2\omega_f-25\omega_Q\approx0.1$. Here, we can also estimate
$\omega_Q$ by measuring the angle $\theta_0$ rotated by the flow
pattern after one forcing period $\tau$, and the result is in fully
agreement with the Poincar\'e section method.

\begin{figure}
\begin{center}
\includegraphics[width=0.7\linewidth]{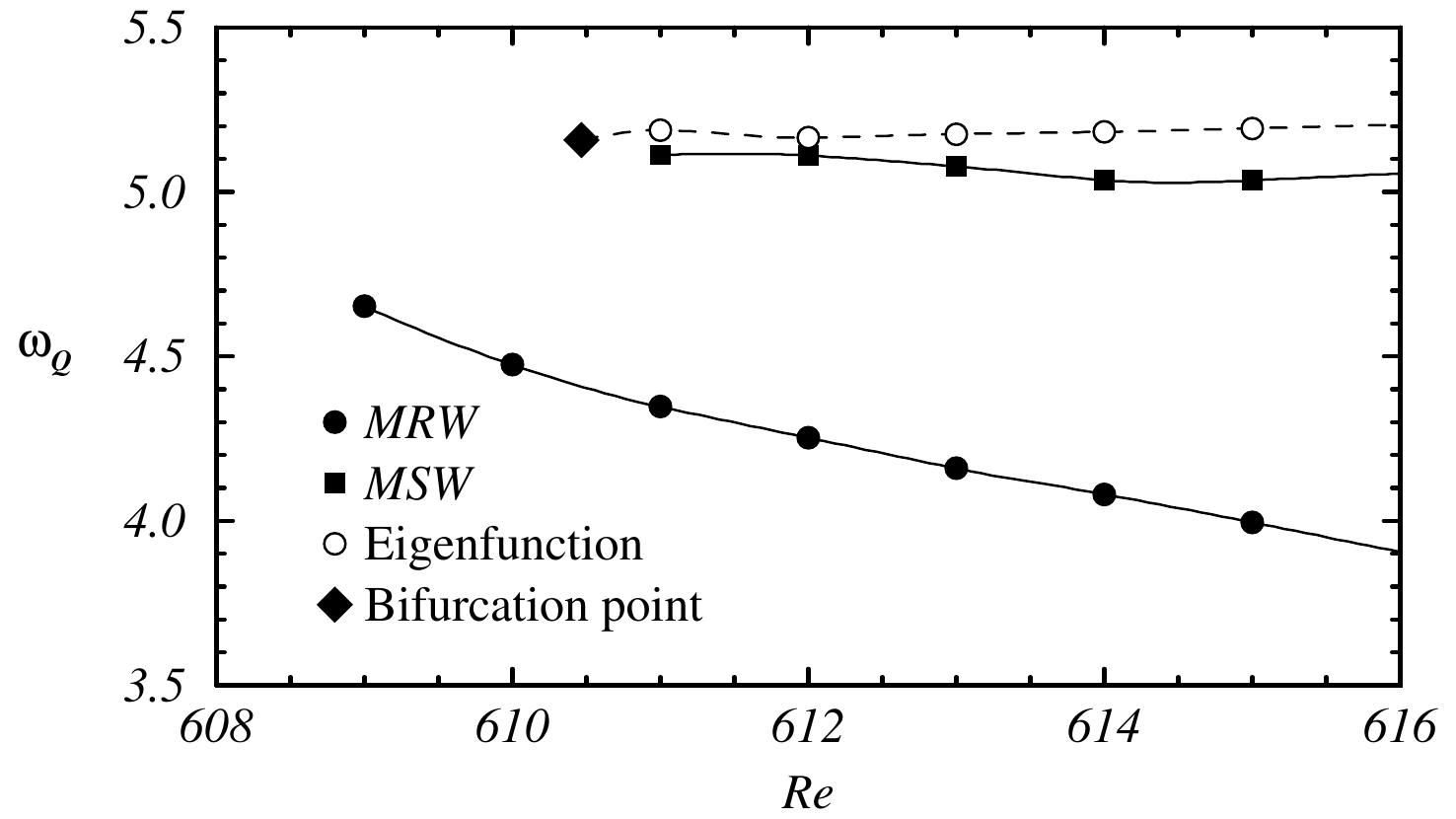}
\end{center}
\caption{Computed frequencies $\omega_Q$ of the quasiperiodic states
  \MRW\ and \MSW\ as a function of $Re$. Also included are the
  frequencies of the most unstable eigenfunction, and the critical
  value $\blacklozenge$.}
\label{QP_frequencies}
\end{figure}

The frequencies of \MRW\ and \MSW\ for $(Re,St)=(615,50)$ are
different, and so we have explored $\omega_Q$ as a function of $Re$,
for fixed $St=50$; the results are shown in
figure~\ref{QP_frequencies}. We have also plotted the critical
frequency at the bifurcation point, $\omega_{Q,\textrm{c}}=5.1572$,
and the value of $\omega_Q$ for the most dangerous eigenfunction as a
function of $Re$. What we observe is that the second frequency for the
eigenfunction is almost constant, as is that for \MSW\ with $\omega_Q$
slightly smaller than the critical frequency
$\omega_{Q,\textrm{c}}$. In contrast, the second frequency of \MRW\ is
substantially smaller than the critical value, and it decreases with
$Re$, the amplitude of the forcing. This is probably related to the
fact that the energy of the \MRW\ is much larger than the energy of
\MSW, and also to the larger subcriticality of the modulated rotating
waves, as shown in figure~\ref{subcrit_QP}.

We see from figure~\ref{QP_frequencies} that the second frequency
$\omega_Q$, for the \MRW\ and \MSW\ at $(Re,St)=(615,50)$, which are
the ones we have discussed in detail in this study, is very close to 4
and 5 respectively. As the forcing frequency is $St=50$, the ratio
$\omega_Q/\omega_f$ is very close to rational (2/25 and 1/10
respectively), as we have already discussed when measuring the
frequencies via FFT and Poincar\'e sections. Of course, along the
curves $\omega_Q(Re)$, other resonances can be located, but all of
them have large denominators, so we do not expect any new dynamics
associated with these resonances \citep{Kuz04}.

\section{Conclusions}\label{conclusions}

Several fluid systems with complete symmetry group $Z_2\times SO(2)$
have been explored in recent years. The interest in these flows was
triggered by the analysis of symmetric bluff body wakes \citep[see a
  summary in][]{BML05}, starting with circular cylinders, and followed
by other symmetric bodies, the square cylinder and a flat plate. Flows
driven by the periodic motion of one of the container walls resulted
in systems with the same symmetry group, and previous studies have
analyzed the rectangular driven cavity \citep{BlLo03_jfm,LHBML05} and
a driven annular cavity \citep{BlLo10}. As the symmetry group of all
these flows is the same, dynamical systems analysis \citep{MLB04}
predicts the type of bifurcations they can undergo, which are the same
in all cases regardless of the specifics of the problem and of the
physical mechanisms at work. There are only three possibilities for
the transition from the basic state to three-dimensional flows:
synchronous modes preserving or breaking the space-time symmetry $H$
or quasiperiodic modes, that come in two flavors, either modulated
travelling waves or modulated standing waves. The synchronous modes
with the corresponding symmetry properties have been observed in all
of these various flow problems. The quasiperiodic modes however have
been much more elusive. In symmetric bluff-body wake problems, the
quasiperiodic modes do not manifest themselves as primary
bifurcations, and can only be observed or computed as secondary or
higher bifurcations, in the form of mixed modes. In periodically
driven flows, as there are more control parameters, they have been
observed in the rectangular driven cavity, but unfortunately spanwise
endwalls effects, effectively breaking the $O(2)$ symmetry, resulted
in modulated travelling waves that do not travel. In the annular
driven geometry with large radius ratios, the quasiperiodic modes are
of very high azimuthal wavenumber and have not been found as
nonlinearly saturated pure modes, but instead they are mixed
complicated modes.

In the present study, we have analyzed the simplest geometry available
with the correct symmetries, the cylindrical driven cavity with
moderate aspect ratio. This geometry has two advantages over previous
studies. First, the $O(2)$ symmetry is exactly fulfilled by the
cylindrical geometry (periodicity in the azimuthal direction),
eliminating the spanwise endwall effects of the rectangular cavity.
And second, the bifurcated states have small azimuthal wavenumbers
(typically $m=1$ or 2), so the competition between different modes is
greatly reduced. As a result, we have been able to compute nonlinearly
saturated pure modulated standing and traveling waves for the first
time, and we have also found the two types of synchronous modes, in
the appropriate parameter ranges.

As a starting point of the analysis, the periodic synchronous base
states have been computed for different forcing amplitudes, $Re$, and
forcing frequencies, $St$. These are non-trivial states, with
axisymmetric rollers forming alternatively close to each of the
endwalls due to the periodic oscillation of the cylinder
sidewall. This oscillation produces axisymmetric jets of azimuthal
vorticity, emerging from the corners, moving into the interior, and 
forming the rollers.

The linear stability analysis of the base state has resulted in the
computation of the marginal stability curve, shown in
figure~\ref{critical}. We have found synchronous bifurcations
preserving the $H$ symmetry for small forcing frequencies $St$, and
breaking $H$ for larger $St$ values. In between, for intermediate $St$
values, we have found a transition to quasiperiodic solutions. The
form of the instabilities is always the same, the formation of braids
that are small-scale meridional perturbations of the rollers. The size
and persistence of these braids depends strongly on $St$. These
results are in good agreement with previous results in flows with the
same symmetries, in particular with the driven rectangular cavity
problem.

We have also computed saturated nonlinear states, and in all cases
sufficiently close to the bifurcation curve, these are pure modes that
we have analyzed in detail. The quasiperiodic stable solutions in the
present problem are modulated rotating waves, and by restricting the
computations to the appropriate subspace, we have also been able to
compute the corresponding unstable modulated standing waves. As a
result of these nonlinear simulations, we have established that the
bifurcations to synchronous states are supercritical, while the
bifurcations to quasiperiodic states are subcritical. A careful
analysis of the quasiperiodic frequency $\omega_Q$ of the modulated
standing waves has shown that $\omega_Q$ is almost constant and close
to the frequency that emerges from the linear stability analysis for
the modulated standing waves, whereas the modulated rotating waves
exhibit a smaller frequency $\omega_Q$ that varies significantly
with $St$. Finally, we have found in preliminary explorations to
higher $Re$ (i.e.\ increasing the forcing amplitude) that the
three-dimensional pure modes undergo secondary bifurcations to
complicated mixed modes for moderate increments in $Re$ beyond
critical.


Future directions include an examination of the dynamics in the
neighborhood of the codimension-two points where two distinct modes
bifurcate simultaneously; these codimension-two points act as
organizing centers of the dynamics, and are very likely associated
with the secondary bifurcations to mixed modes and more complex
dynamics. We have found three codimension-two bifurcations, one
associated with the competition between modes B1 and B2, and two on
each side of the quasiperiodic region. The codimension-two bifurcation
between B1 and B2 is a 1:1 resonance preserving $H$-symmetry, that
has been fully studied theoretically \citep{Kuz04}. The other two
bifurcations are more complex, they are bifurcations of maps with a
real eigenvalue (either $\pm1$) and a pair of complex conjugate
eigenvalues of modulus 1, where the $H$-symmetry plays a key
role. Although there are some partial results on these bifurcations,
they have not been yet fully analyzed \citep{Kuz04}, and here we have
a physically realizable fluid dynamics system in which they
occur. Studies in fluids have been instrumental in developments in
nonlinear dynamics, allowing for unified understanding of complex
dynamics in a wide spectrum of fields. We hope that this driven
cylinder system can help further address some general open questions
in dynamical systems. Moreover, this is a relatively easy system in
which to conduct experimental research, and we hope experiments in
this system will be undertaken in the near future.

\begin{acknowledgments}
  This work was supported by the National Science Foundation grant
  DMS-05052705 and the Spanish Government grant FIS2009-08821. 
\end{acknowledgments}

\bibliography{local}
\bibliographystyle{jfm}

\end{document}